\documentclass[pra,twocolumn,showpacs,preprintnumbers,amsmath,amssymb,superscriptaddress]{revtex4}

\usepackage{graphicx,amsmath,amssymb,amsfonts,latexsym,color,dcolumn,bm,epsfig,textcomp}
\usepackage{epsfig}
\usepackage{bm}
\usepackage{dsfont}
\newcommand{\bra}[1]{\langle#1|}
\newcommand{\ket}[1]{|#1\rangle}

\def\bbm[#1]{\mbox{\boldmath $#1$}}

\usepackage[english]{babel}

\DeclareMathOperator{\Tr}{Tr}
\DeclareMathOperator{\Rea}{Re}
\DeclareMathOperator{\Ima}{Im}
\def\bbm[#1]{\mbox{\boldmath $#1$}}
\newcommand{\TE}{\text{TE}}
\newcommand{\TM}{\text{TM}}

\def\fl{\rightarrow}
\def\e{\varepsilon}

\begin{document}
\title{Quantum systems in a stationary environment out of thermal equilibrium}

\author{Bruno Bellomo}\affiliation{Universit\'{e} Montpellier 2, Laboratoire Charles Coulomb UMR 5221 - F-34095, Montpellier, France}\affiliation{CNRS, Laboratoire Charles Coulomb UMR 5221 - F-34095, Montpellier, France}
\author{Riccardo Messina}\affiliation{Laboratoire Charles Fabry, Institut d'Optique, CNRS, Universit\'{e} Paris-Sud, Campus Polytechnique, RD 128, 91127 Palaiseau cedex, France}
\author{Didier Felbacq}\affiliation{Universit\'{e} Montpellier 2, Laboratoire Charles Coulomb UMR 5221 - F-34095, Montpellier, France}\affiliation{CNRS, Laboratoire Charles Coulomb UMR 5221 - F-34095, Montpellier, France}
\author{Mauro Antezza}\affiliation{Universit\'{e} Montpellier 2, Laboratoire Charles Coulomb UMR 5221 - F-34095, Montpellier, France}\affiliation{CNRS, Laboratoire Charles Coulomb UMR 5221 - F-34095, Montpellier, France}

\date{\today}

\begin{abstract}

We discuss how the thermalization of an elementary quantum system is modified when the system is placed in an environment out of thermal equilibrium. To this aim we provide a detailed investigation of the dynamics of an atomic system placed close to a body of arbitrary geometry and dielectric permittivity, whose temperature $T_\textrm{M}$ is different from that of the surrounding walls $T_\textrm{W}$. A suitable master equation for the general case of an $N$-level atom is first derived and then specialized to the cases of a two- and three-level atom. Transition rates and steady states are explicitly expressed as a function of the scattering matrices of the body and become both qualitatively and quantitatively different from the
case of radiation at thermal equilibrium. Out of equilibrium, the system steady state depends on the system-body distance, on the geometry of the body and on the interplay of all such parameters with the body optical resonances. While a two-level atom tends toward a thermal state, this is not the case already in the presence of three atomic levels. This peculiar behavior can be exploited, for example, to invert the populations ordering and to provide an efficient cooling mechanism for the internal state of the quantum system. We finally provide numerical studies and asymptotic expressions when the body is a slab of finite thickness. Our predictions can be relevant for a wide class of experimental configurations out of thermal equilibrium  involving different physical realizations of two or three-level systems.
\end{abstract}


\maketitle

\section{Introduction}

The absence of thermal equilibrium is a condition naturally verified in several biological systems as well as experimental configurations \cite{CaiPRE10,BrunnerPRE12,Camalet2011}.
Out of thermal equilibrium (OTE) systems have been recently subject to intensive investigations concerning heat transfer \cite{RousseauNP, ShenNL, Ben-AbdallahPRB2010, OttensPRL, MesAntEPL11, MesAntPRA11,KardarPRL,RodriguezPRL,Messina2012} and Casimir-Lifshitz interaction \cite{AntezzaPRL05,  ObrechtPRL07, AntezzaJPA06, AntezzaPRLA06, AntezzaPRA08, BuhmannPRL08, SherkunovPRA09, BehuninPRA2011, MesAntEPL11, MesAntPRA11}. It has been  theoretically and experimentally demonstrated that systems driven out of thermal equilibrium may show forces with new qualitative and quantitative behaviors: repulsion, amplification of the force and stronger tunability \cite{AntezzaPRL05,ObrechtPRL07}. Typical OTE configurations consist of an ensemble of bodies kept at fixed and different temperatures and embedded in a blackbody radiation emitted by the surrounding walls at a temperature different from those of the bodies, the whole system being in a stationary configuration.

Recently, promising results have been also obtained regarding the potentiality of OTE environments to control and manipulate the dynamics of atomic systems placed in this kind of environments \cite{BellomoArxiv12}. This study may have experimental relevance for systems which are naturally out of thermal equilibrium, such as in recent studies concerning cold atoms close to superconducting surfaces \cite{GuerlinNature07,SayrinNature11}, involving the tip of an atomic-force microscope AFM close to cold samples \cite{HenkelJOptA02,JoulainarXiv12}, in the case of atom-chip configurations \cite{ColombeNature07,FortagRevModPhys07} and in all the setups aiming at investigating near-field properties \cite{AntezzaPRL05,KittelPRL05,DeWildeNature06,ObrechtPRL07,KawataNaturePhotonics09}.

The lifetime 1/$\Gamma$ of the atomic dynamics depends on the state of the local electromagnetic (EM) field with which the atom is coupled, i.e. on the fact that it is stationary or not, isotropic or not, that it is a vacuum state, a thermal state, a laser field, or other. Given an atom in a certain position, the local EM field is in general modified by the radiation emitted and/or scattered by surrounding bodies. This mechanism results in lifetimes depending on the relative distance between the atom and the bodies, on their geometrical and optical properties, as well as on their temperatures. Lifetime has been typically studied at thermal equilibrium, when the radiation emitted by the surrounding walls impinging the body is at thermal equilibrium with the body itself. Under this assumption several configurations have been investigated, among which are zero and non-zero temperatures, and infinitely thick planar and corrugated slabs \cite{Novotny06,BiehsPRA11,BenAbdallaharXiv11}.

The atomic lifetime $1/\Gamma$ is strongly connected with the time evolution of atomic coherences, which may be naturally investigated by studying the reduced density matrix, also useful to evaluate the average values of physical observables. Remarkably, it gives information on the atomic steady state and hence on the {\it thermalization process}. For instance, for configurations where the body is at thermal equilibrium with the environment at temperature $T$, the atomic density matrix evolves on a time scale $1/\Gamma$ to a diagonal matrix corresponding to a {\it thermal steady state at temperature} $T$. Contrarily to lifetime, and due to peculiar cancelations, at thermal equilibrium the atomic steady state depends only on the ratios $\hbar\omega_{nm}/k_{\rm B}T$ (where $\hbar\omega_{nm}$ are the energies of the internal transitions), being {\it independent} of the presence of the body. The richness of this system can be
exalted if the atom is embedded in a stationary {\it configuration out of thermal equilibrium}, made by a body whose temperature is kept fixed and different from that of the walls surrounding the body-atom system. The electromagnetic structure holding in OTE systems has a complex nature, poorly investigated because of the impossibility of exploiting directly the fluctuation-dissipation theorem. Recently, using multiple-scattering theory and an appropriate use of the fluctuation-dissipation theorem at each different temperature present in the system, the radiation field in complex OTE configurations has been characterized in terms of its correlators \cite{MesAntEPL11,MesAntPRA11}. The knowledge of the correlators of the local EM field is the main ingredient needed to compute the transition rates regulating the atomic dynamics, as e.g. in the case of Kossakowski-Lindblad master equations in the Markovian regime \cite{BellomoArxiv12,BookBreuer}.

In this paper we provide a systematic derivation for the internal dynamics of an atom placed in front of an arbitrary body embedded in a thermal radiation whose temperature is different from that of the body. The paper is organized as follows. In Sec. \ref{par:model} we describe the physical model under investigation and we derive a master equation for the general case of a $N$-level atom. In Sec. \ref{par:one body} we derive closed-form expressions for the transition rates holding out of thermal equilibrium in terms of the scattering matrices of the body, valid for {\it arbitrary geometrical and material properties}. In Secs. \ref{par:application Two-level system} and \ref{par:three level atom} we specialize our analysis to the case of a two- and a three-level atom, discussing how the atomic dynamics occurring at thermal equilibrium is modified by the absence of equilibrium. In Sec. \ref{par: atom in front of a slab} we derive
explicit expressions for the transition rates for the case in which the body is a slab of finite thickness. This configuration is numerically investigated in Sec. \ref{par:Numerical analysis} where the cases in which slab is made of a dielectrics (GaAs) or a metal (gold) are considered and compared. The dynamics of both two- and three-level atoms are discussed, showing the occurrence of peculiar phenomena emerging out of thermal equilibrium, already pointed out in the general analysis. In Sec. \ref{par:Conclusions} we draw our conclusions.

\section{Model}\label{par:model}

\begin{figure}[b]
\includegraphics[width=0.49\textwidth]{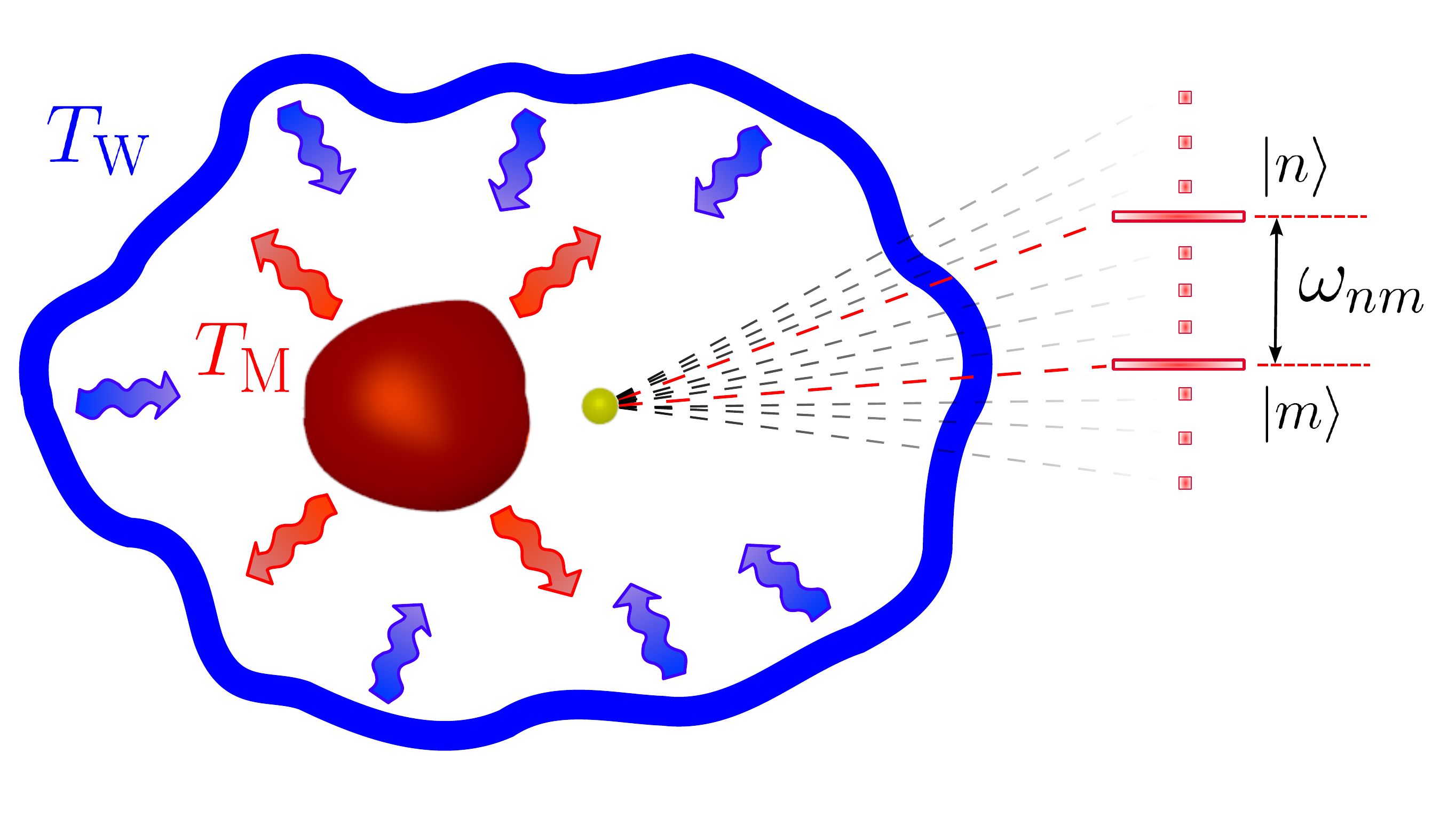}
\caption{\label{fig:1}\footnotesize (color online). The atom is kept fixed close to an arbitrary body whose temperature $T_\mathrm{M}$ is different from that of the surrounding walls, $T_\mathrm{W}$. $T_\mathrm{M}$ and $T_\mathrm{W}$ are kept fixed in time, realizing a stationary environment for the atom.}
\end{figure}

We consider a $N$-level atom $A$ placed close to a body of arbitrary geometry and dielectric permittivity and embedded in an environmental radiation generated by the walls surrounding the system (see Fig. \ref{fig:1}). The body and the surrounding walls have in general different temperatures, $T_\mathrm{M}$ and $T_\mathrm{W}$, which are kept fixed in time realizing a stationary configuration out of thermal equilibrium. By assumption, the walls are far from both the body and the atom, their shape is irregular and they are at local thermal equilibrium at temperature $T_\mathrm{W}$. As a result of these hypotheses, the radiation associated to the wall has no evanescent contribution reaching the body and the atom, and is a universal isotropic blackbody radiation. This blackbody radiation is by definition independent of the actual material constituting the walls. The atom interacts with the electromagnetic field (playing the role of bath $B$) generated by the walls and the body.
The total Hamiltonian has the form
\begin{equation}\label{Hamiltonian}H=H_A+H_B+H_I,\end{equation}
where $H_A$ in the free Hamiltonian of the atom and $H_B$ the free Hamiltonian of the bath. We describe the interaction between the atom and the field using the multipolar coupling in dipole approximation \cite{CohenTannoudji97} $H_I=-\mathbf{D}\cdot\mathbf{E}(\mathbf{R})$, where $\mathbf{D}$ is the atomic electric-dipole operator and $\mathbf{E}(\mathbf{R})$ is the electric field at the atomic position $\mathbf{R}$ in the Schr\"{o}dinger picture.
In the interaction picture $H_I$ becomes
\begin{equation}\label{Interaction hamiltonian}H_I(t)=-\mathbf{D}(t)\cdot\mathbf{E}(\mathbf{R},t),\end{equation}
where the time-dependent electric-dipole operator and electric field are defined by the transformations $\mathbf{D}(t)=\exp(\frac{i}{\hbar}H_At)\mathbf{D}\exp(-\frac{i}{\hbar}H_At)$ and $\mathbf{E}(\mathbf{R},t)=\exp(\frac{i}{\hbar}H_Bt)E(\mathbf{R})\exp(-\frac{i}{\hbar}H_Bt)$. We describe the electric field in the interaction picture using a decomposition in which a mode of the field is identified by the frequency $\omega$, the transverse wave vector $\mathbf{k}=(k_x,k_y)$, the polarization index $p$ (taking the values $p=1,2$ corresponding to TE and TM polarizations respectively), and the direction or propagation $\phi=\pm1$ (shorthand notation $\phi=\pm$) along the $z$ axis. In this approach, the total wavevector takes the form $\mathbf{K}^\phi=(\mathbf{k},\phi k_z)$, where the $z$ component of the wavevector $k_z$ is a dependent variable given by $k_z=\sqrt{\frac{\omega^2}{c^2}-k^2}$, with $k=|\mathbf{k}|$. The explicit expression of the field is
\begin{equation}\label{electric field}\mathbf{E}(\mathbf{R},t)=2\Rea\Biggl[\int_0^{+\infty}\frac{d\omega}{2\pi}e^{-i\omega t}\mathbf{E}(\mathbf{R},\omega)\Biggr],\end{equation}
where a single-frequency component reads
\begin{equation}\label{Eomega}\mathbf{E}(\mathbf{R},\omega)=\sum_{\phi,p}\int\frac{d^2\mathbf{k}}{(2\pi)^2}e^{i\mathbf{K}^\phi\cdot\mathbf{R}}\hat{\bbm[\epsilon]}^\phi_{p}(\mathbf{k},\omega)E^\phi_p(\mathbf{k},\omega),\end{equation}
$E^\phi_p(\mathbf{k},\omega)$ being the field amplitude operator associated to the mode $(\omega, \mathbf{k}, p, \phi)$.
For the polarization vectors appearing in \eqref{Eomega} we adopt the following standard definitions
\begin{equation}\begin{split}
\hat{\bbm[\epsilon]}^\phi_\TE(\mathbf{k},\omega)&=\hat{\mathbf{z}}\times\hat{\mathbf{k}}=\frac{1}{k}(-k_y\hat{\mathbf{x}}+k_x\hat{\mathbf{y}}),\\
\hat{\bbm[\epsilon]}^\phi_\TM(\mathbf{k},\omega)&=\frac{c}{\omega}\hat{\bbm[\epsilon]}^\phi_\TE(\mathbf{k},\omega)\times\mathbf{K}^{\phi}=\frac{c}{\omega}(-k\hat{\mathbf{z}}+\phi k_z\hat{\mathbf{k}}),\end{split}\end{equation}
where $\hat{\mathbf{x}}$, $\hat{\mathbf{y}}$ and $\hat{\mathbf{z}}$ are the unit vectors along the three axes and $\hat{\mathbf{k}}=\mathbf{k}/k$.

\subsection{Derivation of the master equation}\label{par:derivation}

Following \cite{BookBreuer}, we now derive a master equation for the reduced density matrix $\rho(t)=\mathrm{Tr}_B[\rho_{\text{tot}}(t)]$, where $\mathrm{Tr}_B$ denotes the trace over the degrees of freedom associated to the bath. In the following we denote the eigenvalues of $H_A$ by $\epsilon$ and the projection onto the eigenspace belonging to the eigenvalue $\epsilon$ by $\Pi(\epsilon)$.  The starting point of our derivation is, in the interaction picture, the von Neumann equation for the total density matrix $\rho_{\text{tot}}(t)$:
\begin{equation}\label{von Neummann equation}\frac{d}{dt}\rho_{\text{tot}}(t)=-\frac{i}{\hbar}[H_I(t),\rho_{\text{tot}}(t)],\end{equation}
where $H_I(t)$ of Eq. \eqref{Interaction hamiltonian} is rewritten as \cite{BookBreuer}
\begin{equation}\label{Interaction hamiltonian 2}H_I(t)=-\sum_{i,\omega}e^{-i\omega t} A_{i}(\omega) E_i(\mathbf{R},t),\end{equation}
where $i\, \in\, \{x, y, z\}$ and $A_{i}(\omega)= \sum_{\epsilon'-\epsilon=\omega}\Pi(\epsilon)D_i\Pi(\epsilon')$. $A_{i}(\omega)$ are eigenoperators of $H_A$ belonging to the frequencies $\pm \omega$, being $[H_A,A_i(\omega)]=-\omega A_i(\omega)$ and $[H_A,A_i^\dag(\omega)]=\omega A_i^\dag(\omega)$, and it holds $D_i=\sum_\omega A_i(\omega)$ and $A_i(-\omega)=A_i^\dag (\omega)$. The sum over $\omega$ appearing in \eqref{Interaction hamiltonian 2} is extended over all the energy eigenvalues $\epsilon'$ and $\epsilon$ of $H_A$ such that  $\epsilon'-\epsilon=\omega$.

Using standard approximations, namely the Born, Markovian and rotating-wave approximations (see discussion in Appendix \ref{par:discussions on approximations}), one obtains (using also the condition $\langle E_i(\mathbf{R},t)\rangle=0$):
\begin{equation}\label{master equation}\frac{d}{d t}\rho(t)=-\frac{i}{\hbar}[H_{LS},\rho(t)]+\mathcal{D}\bigl(\rho(t)\bigr),\end{equation}
where the Hermitian operator (Lamb-shift Hamiltonian)
\begin{equation}\label{Lamb shift Hamiltonian}H_{LS}=\hbar\sum_\omega\sum_{i,j} s_{i j}(\omega)A^\dag_i(\omega)A_j(\omega)\end{equation}
purely produces a shift of the atomic energy levels (we note that $[H_A,H_{LS}]=0$) and the dissipator term $\mathcal{D}\bigl(\rho(t)\bigr)$ has the form
\begin{equation}\label{dissipator term}\begin{split}\mathcal{D}\bigl(\rho(t)\bigr)&=\sum_\omega\sum_{i,j}\gamma_{ij}(\omega)\Big(A_j(\omega)\rho(t)A^\dag_i(\omega)\\
&\hspace{2cm}-\frac{1}{2}\{A^\dag_i(\omega)A_j(\omega),\rho(t)\}\Big),\end{split}\end{equation}
where in Eqs. \eqref{Lamb shift Hamiltonian} and \eqref{dissipator term} the frequencies $\omega$ assume both positive and negative values.
 In the two previous equations, $\gamma_{ij}(\omega)$ and $s_{ij}(\omega)$ are defined by
\begin{equation}\label{Xi function}\begin{split} 
\Sigma_{ij}(\omega)&= \frac{1}{\hbar^2}\int_0^\infty  \!\!ds\,e^{i\omega s}\langle E_i(\mathbf{R},t)E_{j}(\mathbf{R},t-s)\rangle , \\
\gamma_{ij}(\omega) &=\Sigma_{ij} +\Sigma_{ji}^{*} ,\quad
s_{ij}(\omega) =  \frac{\Sigma_{ij} -\Sigma_{j i }^{*}}{2 i} .
\end{split}\end{equation}
From the previous equation it follows that $\Sigma_{ij}(\omega)=\frac{1}{2}\gamma_{ij}(\omega)+i s_{ij}(\omega)$ and
\begin{equation}\label{gammaijdef}\gamma_{i j}(\omega)=\frac{1}{\hbar^2}\int_{-\infty}^\infty ds\,e^{i\omega s}\langle E_i(\mathbf{R},t)E_j(\mathbf{R},t-s)\rangle.\end{equation}
The quantities $\gamma_{i j}(\omega)$ are expressed in terms of the reservoir correlations functions and in the case $\rho_B$ is a stationary state of the bath (as it is in our physical configuration), that is $[H_B,\rho_B]=0$, they are homogenous in time. This means that $\langle E_i(\mathbf{R},t)E_j(\mathbf{R},t-s)\rangle=\langle E_i(\mathbf{R},s)E_j(\mathbf{R},0)\rangle$ and $\gamma_{i j}(\omega)$ do not depend on time.

Field correlators have been computed out of thermal equilibrium in  a more general configuration, in presence of a further object at a third different temperature \cite{MesAntEPL11,MesAntPRA11}. These depend on the three temperatures involved as well as on the material and geometrical properties of the two bodies, taken into account by means of their classical scattering operators. We will show that by exploiting these results, the expressions of the transition rates can be explicitly given as a function of the reflection and transmission operators $\mathcal{R}$ and $\mathcal{T}$ of the body M.

\subsection{$N$-level atom}\label{par:N level atom}

Now we explicitly apply the previous derivation to an atomic system. We first consider the general case of an atom having $N$ states (ordered by increasing energy) $\ket{1},\ket{2}, ...$ and $\ket{N}$, with corresponding frequencies $\omega_1,\omega_2,\dots,\omega_N$. All the transitions between the $N$ states are in principle allowed and the frequency difference between two arbitrary levels $n$ and $m$ is indicated by $\omega_{nm}=\omega_n-\omega_m$. We will assume from now on that $n$ is always taken larger than $m$ so that $\omega_{nm}$ always represents a positive frequency. The free Hamiltonian of the $N$-level atom is
\begin{equation}\label{Atom hamiltonian N}H_A=\sum_{n=1}^N\hbar\omega_n\ket{n}\bra{n},\end{equation}
and the atomic dipole operator $\mathbf{D}(t)$ takes the form
\begin{equation}\label{dipole operator N}\mathbf{D}(t)=\sum_{m,n}\Bigl(\mathbf{d}_{mn}\ket{m}\bra{n}e^{-i\omega_{nm}t}+\mathbf{d}^*_{mn}\ket{n}\bra{m}e^{i\omega_{nm} t}\Bigr),\end{equation}
where $\mathbf{d}_{mn}=\bra{m}\mathbf{D}\ket{n}$ is the transition matrix element of the dipole operator between $\ket{m}$ and $\ket{n}$. In accordance with the previous remark, the sum runs over values $m,n\in\{1,2,\dots,N\}$ such that $n>m$. Under these assumptions, the sum over $\omega$ in Eqs. \eqref{Interaction hamiltonian 2}, \eqref{Lamb shift Hamiltonian} and \eqref{dissipator term} runs over the values $\omega_{nm}$ and $-\omega_{nm}$. By comparing Eqs. \eqref{Interaction hamiltonian}, \eqref{Interaction hamiltonian 2} and \eqref{dipole operator N} one sees that  (we remark that from now on $\omega>0$)
\begin{equation}\label{A general form} 
\mathbf{A}(\omega)= \sum_{\{m,n\}: \omega_{nm}=\omega} \mathbf{d}_{mn}\ket{m}\bra{n}=\mathbf{A}^\dag(-\omega),
\end{equation}
meaning that the sum is over $n$ and $m$ such that $\omega_{nm}=\omega$. In general several transitions can be associated to the same atomic frequency $\omega$, both because of degeneracy and/or the occurrence of equidistant levels. By inserting Eq. \eqref{A general form} in Eq. \eqref{master equation} and coming back to the Schr\"{o}dinger representation, one obtains the master equation in its most general form 
\begin{equation}\label{master equation N general }\begin{split}&\frac{d}{dt}\rho(t)=
-i\Biggl[\sum_n{\omega_n}\ket{n}\bra{n}+   \sum_{\omega} \sum_{\{m,n\}, \{m',n'\}} \sum_{i,j}  \\
&\Big(s_{ij}(-\omega) [\textbf{d}_{m'n'}]_{i}[\textbf{d}_{mn}]^*_{j} \delta_{nn'}\ket{m'}\bra{m}\\
&+ s_{ij}(\omega)[\textbf{d}_{mn}]^*_{i} [\textbf{d}_{m'n'}]_{j} \delta_{mm'} \ket{n}\bra{n'}\Big),\rho(t)\Biggr]\\
&+\sum_{\omega}\sum_{\{m,n\},\{m',n'\}} \sum_{i,j}\Biggl[ \gamma_{i j}(-\omega)[\textbf{d}_{m'n'}]_{i}[\textbf{d}_{mn}]^*_{j}\\
&\times\Big(\rho_{mm'}\ket{n}\bra{n'}-\frac{1}{2}\delta_{nn'}\{\ket{m'}\bra{m},\rho(t)\}\Big)\\
&+\gamma_{i j}(\omega)[\textbf{d}_{mn}]^*_{i} [\textbf{d}_{m'n'}]_{j}\\
&\times \Big(\rho_{n'n}\ket{m'}\bra{m}-\frac{1}{2}\delta_{mm'}\{\ket{n}\bra{n'},\rho(t)\}\Big)\Biggr],
\end{split}\end{equation}
where the sum $\sum_{\{m,n\},\{m',n'\}}$ is meant over all couples $(m,n)$ and $(m',n')$ such that $\omega_{nm}=\omega_{n'm'}=\omega$ (being alway $n>m$ and $n'>m'$).
The previous equation contains both terms coupling each transition with itself ($(m,n)=(m',n')$) and terms coupling different transitions having the same frequency  ($(m,n)\neq(m',n')$).
In the case of non-equidistant and non-degenerate levels, we have that  to each transition corresponds a different value $\omega_{nm}$ and we can write
$\mathbf{A}({\omega_{nm}})=\mathbf{d}_{mn}\ket{m}\bra{n}=\mathbf{A}^\dag(-\omega_{nm})$. In this case, the previous master equation reduces to 
\begin{equation}\label{master equation N}\begin{split}&\frac{d}{dt}\rho(t)=-i\Bigl[\sum_n{\omega_n}\ket{n}\bra{n}+\sum_{m,n}S(-\omega_{nm})\ket{m}\bra{m}\\
&+\sum_{m,n} S(\omega_{nm})\ket{n}\bra{n},\rho(t)\Bigr]\\
&+\sum_{m,n}\Gamma(-\omega_{nm})\Big(\rho_{mm}\ket{n}\bra{n}-\frac{1}{2}\{\ket{m}\bra{m},\rho(t)\}\Big)\\
&+\sum_{m,n}\Gamma(\omega_{nm})\Big(\rho_{nn}\ket{m}\bra{m}-\frac{1}{2}\{\ket{n}\bra{n},\rho(t)\}\Big),\end{split}\end{equation}
where we have defined
\begin{equation}\label{me parameters N}\begin{split}S(-\omega_{nm})&=\sum_{i,j}s_{ij}(-\omega_{nm})[\textbf{d}_{mn}]_{i}[\textbf{d}_{mn}]^*_{j},\\
S(\omega_{nm})&=\sum_{i,j} s_{ij}(\omega_{nm})[\textbf{d}_{mn}]^*_{i} [\textbf{d}_{mn}]_{j},\\
\Gamma(-\omega_{nm})&=\sum_{i,j}\gamma_{i j}(-\omega_{nm})[\textbf{d}_{mn}]_{i}[\textbf{d}_{mn}]^*_{j},\\
\Gamma(\omega_{nm})&=\sum_{i,j}\gamma_{i j}(\omega_{nm})[\textbf{d}_{mn}]^*_{i} [\textbf{d}_{mn}]_{j}.\end{split}\end{equation}

We now turn to the calculation of the functions $\gamma_{i j}(\omega)$ appearing in Eq. \eqref{me parameters N} and defined in \eqref{gammaijdef}. They are, using homogeneity in time,
\begin{equation}\label{gammaij}\begin{split}\gamma_{i j}&(\omega)=\frac{1}{\hbar^2}\int_{-\infty}^\infty ds\int_0^{+\infty}\frac{d\omega'}{2\pi}\int_0^{+\infty}\frac{d\omega''}{2\pi}\\
&\,\times\Bigl[e^{-i(\omega''-\omega)s}\langle E_i(\mathbf{R},\omega'')E_j^\dag(\mathbf{R},\omega')\rangle\\
&\,+e^{i(\omega+\omega'')s}\langle E_i^\dag(\mathbf{R},\omega'')E_j(\mathbf{R},\omega')\rangle\Bigr]\\
&=\frac{1}{\hbar^2}\int_0^{+\infty}\frac{ d\omega'}{2\pi}\begin{cases}\langle E_i(\mathbf{R},\omega)E_j^\dag(\mathbf{R},\omega')\rangle & \omega>0\\
\langle E_i^\dag(\mathbf{R},-\omega)E_j(\mathbf{R},\omega')\rangle & \omega<0\end{cases}\end{split}\end{equation}
where we have used $\langle E_i(\mathbf{R},\omega)E_j(\mathbf{R},\omega')\rangle=\langle E_i^\dag(\mathbf{R},\omega)E_j^\dag(\mathbf{R},\omega')\rangle=0$ and $\int_{-\infty}^\infty ds\exp(-i\epsilon s)=2\pi\delta(\epsilon)$. By using the decomposition of Eq. \eqref{Eomega}, we obtain for positive arguments of $\gamma_{ij}(\omega)$
\begin{equation}\label{CorrFreq}\begin{split}&\gamma_{i j}(\omega)=\frac{1}{\hbar^2}\int_0^{+\infty}\frac{d\omega'}{2\pi}\sum_{\phi,\phi',p,p'}\int\frac{d^2\mathbf{k}}{(2\pi)^2}\int\frac{d^2\mathbf{k}'}{(2\pi)^2}
\\& \times e^{i(\mathbf{K}^\phi-\mathbf{K'}^{\phi'*})\cdot\mathbf{R}}[\hat{\bbm[\epsilon]}_p^\phi(\mathbf{k},\omega)]_i[\hat{\bbm[\epsilon]}_{p'}^{\phi'}(\mathbf{k}',\omega')]_j^*\\
&\times\langle E_p^\phi(\mathbf{k},\omega)E_{p'}^{\phi'\dag}(\mathbf{k}',\omega')\rangle , \end  {split}\end{equation}
whereas for negative arguments we have
\begin{equation}\label{CorrFreq-}\begin{split}&\gamma_{i j}(-\omega)=\frac{1}{\hbar^2}\int_0^{+\infty}\frac{ d\omega'}{2\pi}\sum_{\phi,\phi',p,p'}\int\frac{d^2\mathbf{k}}{(2\pi)^2}\int\frac{d^2\mathbf{k}'}{(2\pi)^2}
\\&\times e^{-i (\mathbf{K}^{\phi*}-\mathbf{K'}^{\phi'})\cdot\mathbf{R}}[\hat{\bbm[\epsilon]}_p^\phi(\mathbf{k},\omega)]_i^*[\hat{\bbm[\epsilon]}_{p'}^{\phi'}(\mathbf{k}',\omega')]_j\\ &
\times\langle E_p^{\phi\dag}(\mathbf{k},\omega)E_{p'}^{\phi'}(\mathbf{k}',\omega')\rangle,\end{split}\end{equation}
where we always assume $\omega>0$. We observe that last equation can be obtained by taking the complex conjugate of Eq. \eqref{CorrFreq} after having interchanged the operators $E_p^\phi(\mathbf{k},\omega)$ and $E_{p'}^{\phi'\dag}(\mathbf{k}',\omega')$.

\section{Atom close to an arbitrary body}\label{par:one body}

At this point of the calculation, it is necessary to link the field correlation functions $\langle E_p^\phi(\mathbf{k},\omega)E_{p'}^{\phi'\dag}(\mathbf{k}',\omega')\rangle$ and $\langle E_p^{\phi\dag}(\mathbf{k},\omega)E_{p'}^{\phi'}(\mathbf{k}',\omega')\rangle$ to the two temperatures $T_\text{M}$ and $T_\text{W}$ and to the material and geometrical properties of the body as well. This can be done by following the derivation discussed in \cite{MesAntPRA11} in the more general case of two bodies and three temperatures. We  are interested in the correlation functions of the field defined on the side of the body on which the atom is located (the right side, in our case). For a given set $(\omega,\mathbf{k},p)$, the expression of the the modes of the field propagating in the two directions is straightforward. We have, making the dependence on $\omega$, $\mathbf{k}$ and $p$ implicit,
\begin{equation}\label{totalfield}\begin{cases}E^+=E^{(\text{M})+}+\mathcal{T}E^{\text{(W)}+}+\mathcal{R}E^{\text{\text{(W)}}-}\\
E^-=E^{\text{(W)}-}\end{cases}\end{equation}
The field $E^-$ propagating toward the body (i.e. toward the left) is only the environmental one, while the field $E^+$ propagating toward the right  (see Fig. \eqref{fig:1}) results from the field $E^{(\text{M})+}$ directly produced by the body, the transmission through M of the environmental field coming from the left, and the reflection by M of the one coming from the right. The operators $\mathcal{R}$ and $\mathcal{T}$ are the standard reflection and transmission scattering operators, explicitly defined for example in \cite{MesAntPRA11}, associated in this case to the right side of the body. They connect any outgoing (reflected or transmitted) mode of the field to the entire set of incoming modes.

Total field correlators are obtained by the knowledge of correlators of the fields emitted by each source together with Eq. \eqref{totalfield}. The source fields have been characterized in \cite{MesAntPRA11} by treating each source independently as if it was at thermal equilibrium at its own temperature and thus applying the fluctuation-dissipation theorem. The following symmetrized correlation functions ($\langle AB\rangle_\text{sym}=(\langle AB\rangle+\langle BA\rangle)/2$) have been obtained
\begin{equation}\label{sourcefields}\begin{split}&\langle E^{\text{(M)}+}_p(\mathbf{k},\omega)E^{\text{(M)}+\dag}_{p'}(\mathbf{k}',\omega')\rangle_\text{sym}=\frac{\omega}{2\epsilon_0c^2}N(\omega,T_\textrm{M})\\
&\,\times2\pi\delta(\omega-\omega')\bra{p,\mathbf{k}}\Bigl(\mathcal{P}_{-1}^\text{(pw)}-\mathcal{R}\mathcal{P}_{-1}^\text{(pw)}\mathcal{R}^{\dag}+\mathcal{R}\mathcal{P}_{-1}^\text{(ew)}\\
&\,-\mathcal{P}_{-1}^\text{(ew)}\mathcal{R}^{\dag}-\mathcal{T}\mathcal{P}_{-1}^\text{(pw)}\mathcal{T}^{\dag}\Bigr)\ket{p',\mathbf{k}'},\\
&\langle E^{\text{(W)}\phi}_p(\mathbf{k},\omega)E^{\text{(W)}\phi'\dag}_{p'}(\mathbf{k}',\omega')\rangle_\text{sym}=\frac{\omega}{2\epsilon_0c^2}N(\omega,T_\textrm{W})\\
&\,\times2\pi\delta(\omega-\omega')\delta_{\phi,\phi'}\bra{p,\mathbf{k}}\mathcal{P}_{-1}^{\text{(pw)}}\ket{p',\mathbf{k}'},\\\end{split}\end{equation}
where
\begin{equation}N(\omega,T)=\frac{\hbar\omega}{2}\coth\Bigl(\frac{\hbar\omega}{2k_BT}\Bigr)=\hbar\omega\Bigl[\frac{1}{2}+n(\omega,T)\Bigr],\end{equation}
with
\begin{equation}n(\omega,T)=\Bigl(e^{\frac{\hbar\omega}{k_BT}}-1\Bigr)^{-1},\end{equation}
and
\begin{equation}\bra{p,\mathbf{k}}\mathcal{P}_n^\text{(pw/ew)}\ket{p',\mathbf{k}'}=k_z^n\bra{p,\mathbf{k}}\Pi^\text{(pw/ew)}\ket{p',\mathbf{k}'},\end{equation}
with $\Pi^\text{(pw)}$ and $\Pi^\text{(ew)}$ being the projectors on the propagative ($ck<\omega$, corresponding to a real $k_z$) and evanescent ($ck>\omega$, corresponding to a purely imaginary $k_z$) sectors respectively. By combining Eqs. \eqref{totalfield} and \eqref{sourcefields} we finally obtain the symmetrized correlation functions of the total field in the region of interest
\begin{equation}\label{explicitCorrB2}
\begin{split}&\langle E^{+}_p(\mathbf{k},\omega)E^{+\dag}_{p'}(\mathbf{k}',\omega')\rangle_\text{sym}=2\pi\delta(\omega-\omega')\frac{\omega}{2\epsilon_0c^2}\\
&\,\bra{p,\mathbf{k}}\Bigl[N(\omega,T_\textrm{M})\Bigl(\mathcal{P}_{-1}^\text{(pw)}-\mathcal{R}\mathcal{P}_{-1}^\text{(pw)}\mathcal{R}^{\dag}+\mathcal{R}\mathcal{P}_{-1}^\text{(ew)}\\
&\,-\mathcal{P}_{-1}^\text{(ew)}\mathcal{R}^{\dag}-\mathcal{T}\mathcal{P}_{-1}^\text{(pw)}\mathcal{T}^{\dag}\Bigr)\\
&\,+N(\omega,T_\textrm{W})\Bigl(\mathcal{T}\mathcal{P}_{-1}^{\text{(pw)}}\mathcal{T}^{\dag}+\mathcal{R}\mathcal{P}_{-1}^{\text{(pw)}}\mathcal{R}^{\dag}\Bigr)\Bigr]\ket{p',\mathbf{k}'},\\
&\langle E^{+}_p(\mathbf{k},\omega)E^{-\dag}_{p'}(\mathbf{k}',\omega')\rangle_\text{sym}=2\pi\delta(\omega-\omega')\frac{\omega}{2\epsilon_0c^2} N(\omega,T_\textrm{W})\\&\times\bra{p,\mathbf{k}}\mathcal{R}\mathcal{P}_{-1}^{\text{(pw)}}\ket{p',\mathbf{k}'},\\
&\langle E^{-}_p(\mathbf{k},\omega)E^{+\dag}_{p'}(\mathbf{k}',\omega')\rangle_\text{sym}=2\pi\delta(\omega-\omega')\frac{\omega}{2\epsilon_0c^2}N(\omega,T_\textrm{W})\\&\times\bra{p,\mathbf{k}}\mathcal{P}_{-1}^{\text{(pw)}}\mathcal{R}^{\dag}\ket{p',\mathbf{k}'},\\
&\langle E^{-}_p(\mathbf{k},\omega)E^{-\dag}_{p'}(\mathbf{k}',\omega')\rangle_\text{sym}=2\pi\delta(\omega-\omega')\frac{\omega}{2\epsilon_0c^2}N(\omega,T_\textrm{W})\\&
\times\bra{p,\mathbf{k}}\mathcal{P}_{-1}^{\text{(pw)}}\ket{p',\mathbf{k}'}.\end{split}\end{equation}
To obtain the non-symmetrized versions of these correlation functions, appearing in Eqs. \eqref{CorrFreq} and \eqref{CorrFreq-}, we first remark that the source correlation functions \eqref{sourcefields} have been derived using thermal-equilibrium techniques at the temperature of each source individually (see \cite{MesAntPRA11} for a detailed discussion). As a consequence, we are allowed to use Kubo's prescription \cite{KuboRepProgPhys66}, according to which in order to obtain $\langle AB\rangle$ from $\langle AB\rangle_\text{sym}$ the replacement $N(\omega,T_i)\rightarrow\hbar\omega\bigl[1+n(\omega,T_i)\bigr]$ must be performed, whilst $\langle BA\rangle$ results from the replacement $N(\omega,T_i)\rightarrow\hbar\omega\,n(\omega,T_i)$. By exploiting the former replacement rule, we obtain the modified version of Eq. \eqref{CorrFreq} for $\omega>0$
\begin{equation}\label{gammagenerale}\begin{split}
&\gamma_{i j}(\omega)=\frac{\omega^2}{2\epsilon_0\hbar c^2}\sum_{p,p'}\int\frac{d^2\mathbf{k}}{(2\pi)^2}\int\frac{d^2\mathbf{k}'}{(2\pi)^2}e^{i(\mathbf{k}-\mathbf{k}')\cdot\mathbf{r}}\\
&\,\times\bra{p,\mathbf{k}}\Bigl\{e^{i(k_z-k_z^{'*})z}[\hat{\bbm[\epsilon]}_p^+(\mathbf{k},\omega)]_i[\hat{\bbm[\epsilon]}_{p'}^{+}(\mathbf{k}',\omega)]_j^*\\
&\,\times\Bigl[\bigl[1+n(\omega,T_\textrm{M})\bigr]\Bigl(\mathcal{P}_{-1}^\text{(pw)}-\mathcal{R}\mathcal{P}_{-1}^\text{(pw)}\mathcal{R}^{\dag}\\
&\,+\mathcal{R}\mathcal{P}_{-1}^\text{(ew)}-\mathcal{P}_{-1}^\text{(ew)}\mathcal{R}^{\dag}-\mathcal{T}\mathcal{P}_{-1}^\text{(pw)}\mathcal{T}^{\dag}\Bigr)\\
&\,+\bigl[1+n(\omega,T_\textrm{W})\bigr]\Bigl(\mathcal{T}\mathcal{P}_{-1}^{\text{(pw)}}\mathcal{T}^{\dag}+\mathcal{R}\mathcal{P}_{-1}^{\text{(pw)}}\mathcal{R}^{\dag}\Bigr)\Bigr]\\
&\,+\bigl[1+n(\omega ,T_\textrm{W})\bigr]\Bigl[e^{i(k_z+k_z^{'*})z}[\hat{\bbm[\epsilon]}_p^+(\mathbf{k},\omega)]_i[\hat{\bbm[\epsilon]}_{p'}^{-}(\mathbf{k}',\omega)]_j^*\mathcal{R}\mathcal{P}_{-1}^{\text{(pw)}}\\
&\,+e^{-i(k_z+k_z^{'*})z}[\hat{\bbm[\epsilon]}_p^-(\mathbf{k},\omega)]_i[\hat{\bbm[\epsilon]}_{p'}^{+}(\mathbf{k}',\omega)]_j^*\mathcal{P}_{-1}^{\text{(pw)}}\mathcal{R}^{\dag}\\
&+e^{-i(k_z-k_z^{'*})z}[\hat{\bbm[\epsilon]}_p^-(\mathbf{k},\omega)]_i[\hat{\bbm[\epsilon]}_{p'}^{-}(\mathbf{k}',\omega)]_j^*\mathcal{P}_{-1}^{\text{(pw)}}\Bigr]\Big\}\ket{p',\mathbf{k}'}.\end{split}\end{equation}
We observe that $\gamma_{i j}(-\omega)$ of Eq. \eqref{CorrFreq-} is obtained from the previous equation by replacing $\bigl[1+n(\omega,T_i)\bigr]$ with $n(\omega,T_i)$ and by taking the complex conjugate.

Using Eq. \eqref{gammagenerale} one can finally cast the transition rates $\Gamma(\omega_{nm})$ and $\Gamma(-\omega_{nm})$ of Eq. \eqref{me parameters N} under the form
\begin{equation}\label{GammaN}\begin{split}\begin{pmatrix}\Gamma(\omega_{nm})\\\Gamma(-\omega_{nm})\end{pmatrix}&=\Gamma_0(\omega_{nm})\big[\alpha_\text{W}(\omega_{nm})+\alpha_\text{M}(\omega_{nm})\bigr]\\
&\,\times\begin{pmatrix}1+n_\text{eff}^{(nm)}\\n_\text{eff}^{(nm)}\end{pmatrix} ,\end{split}\end{equation}
where $\Gamma_0(\omega_{nm})=\frac{\omega_{nm}^3|\mathbf{d}_{mn}|^2}{3\pi\epsilon_0\hbar c^3}$ is the vacuum spontaneous-emission rate relative to the transition $\ket{m}\rightleftharpoons\ket{n}$. Here, we have introduced for an arbitrary transition frequency $\omega=\omega_{nm}$
\begin{equation}\label{alphas}\begin{split}&\alpha_\text{W}(\omega)=\frac{3\pi c}{2\omega}\!\sum_{p,p'}\sum_{i,j}\frac{[\mathbf{d}_{mn}]_i^*[\mathbf{d}_{mn}]_j}{|\mathbf{d}_{mn}|^2}\int\frac{d^2\mathbf{k}}{(2\pi)^2}\int\frac{d^2\mathbf{k}'}{(2\pi)^2}\\
&\,e^{i(\mathbf{k}-\mathbf{k'})\cdot\mathbf{r}}\bra{p,\mathbf{k}}\Bigl[e^{-i(k_z-k_z^{'*})z}[\hat{\bbm[\epsilon]}_p^-(\mathbf{k},\omega)]_i[\hat{\bbm[\epsilon]}_{p'}^-(\mathbf{k}',\omega)]_j^*\mathcal{P}_{-1}^{\text{(pw)}}\\
&\,+e^{i(k_z+k_z^{'*})z}[\hat{\bbm[\epsilon]}_p^+(\mathbf{k},\omega)]_i[\hat{\bbm[\epsilon]}_{p'}^{-}(\mathbf{k}',\omega)]_j^*\mathcal{R}\mathcal{P}_{-1}^{\text{(pw)}}\\
&\,+e^{-i(k_z+k_z^{'*})z}[\hat{\bbm[\epsilon]}_p^-(\mathbf{k},\omega)]_i[\hat{\bbm[\epsilon]}_{p'}^{+}(\mathbf{k}',\omega)]_j^*\mathcal{P}_{-1}^{\text{(pw)}}\mathcal{R}^\dag\\
&\,+e^{i(k_z-k_z^{'*})z}[\hat{\bbm[\epsilon]}_p^+(\mathbf{k},\omega)]_i[\hat{\bbm[\epsilon]}_{p'}^{+}(\mathbf{k}',\omega)]_j^*\\
&\,\times\Bigl(\mathcal{T}\mathcal{P}_{-1}^{\text{(pw)}}\mathcal{T}^\dag+\mathcal{R}\mathcal{P}_{-1}^{\text{(pw)}}\mathcal{R}^\dag\Bigr)\Bigr]\ket{p',\mathbf{k}'},\\
&\alpha_\text{M}(\omega)=\frac{3\pi c}{2\omega}\!\sum_{p,p'}\sum_{i,j}\frac{[\mathbf{d}_{mn}]_i^*[\mathbf{d}_{mn}]_j}{|\mathbf{d}_{mn}|^2}\int\frac{d^2\mathbf{k}}{(2\pi)^2}\int\frac{d^2\mathbf{k}'}{(2\pi)^2}\\ &\,e^{i(\mathbf{k}-\mathbf{k}')\cdot\mathbf{r}}\bra{p,\mathbf{k}}e^{i(k_z-k_z^{'*})z}[\hat{\bbm[\epsilon]}_p^+(\mathbf{k},\omega)]_i[\hat{\bbm[\epsilon]}_{p'}^{+}(\mathbf{k}',\omega)]_j^*\Bigl(\mathcal{P}_{-1}^\text{(pw)}\\
&\,-\mathcal{R}\mathcal{P}_{-1}^\text{(pw)}\mathcal{R}^\dag+\mathcal{R}\mathcal{P}_{-1}^\text{(ew)}-\mathcal{P}_{-1}^\text{(ew)}\mathcal{R}^\dag-\mathcal{T}\mathcal{P}_{-1}^\text{(pw)}\mathcal{T}^\dag\Bigr)\ket{p',\mathbf{k}'}\end{split}\end{equation}
and
\begin{equation}\label{n effective ij}n_{\mathrm{eff}}^{(nm)}=\frac{n(\omega_{nm},T_\mathrm{W})\alpha_\mathrm{W}(\omega_{nm})+n(\omega_{nm},T_\mathrm{M})\alpha_\mathrm{M}(\omega_{nm})}{\alpha_\mathrm{W}(\omega_{nm})+\alpha_\mathrm{M}(\omega_{nm})}.\end{equation}
The transition rates $\Gamma(\pm\omega_{nm})$ of Eq. \eqref{GammaN} are proportional to the term $\alpha_\text{W}(\omega_{nm})+\alpha_\text{M}(\omega_{nm})$ which depends on material properties but not on temperatures, while $n_{\mathrm{eff}}^{(nm)}$ depends, in general, both on temperatures and the properties of the body. From the last equation it follows that $n(\omega_{nm},T_\text{min})<n_\text{eff}^{(nm)}<n(\omega_{nm},T_\text{max})$, where $T_\text{min}=\min(T_\text{M},T_\text{W})$ and $T_\text{max}=\max(T_\text{M},T_\text{W})$. As a consequence, the values of the transition rates $\Gamma(\pm\omega_{nm})$ of Eq. \eqref{GammaN} are always confined between their equilibrium values at temperatures $T_\text{M}$ and $T_\text{W}$. The transition rates relative to each frequency $\omega_{nm}$ are equal to the ones at thermal equilibrium at an effective temperature $T_{\mathrm{eff}}^{({nm})}$, associable to each transition. This is defined by $n(\omega_{nm},T_{\mathrm{eff}}^{({nm})})=n_{\mathrm{eff}}^{({
nm})}$
and is given by
\begin{equation}\label{effective temperature ij}T_{\mathrm{eff}}^{({nm})}=\frac{\hbar\omega_{{nm}}}{k_B }\Bigl[\ln\Bigl(1+{n_{\mathrm{eff}}^{({nm})}}^{-1}\Bigr)\Bigr]^{-1},\end{equation}
being in general $T_{\mathrm{eff}}^{(nm)}\neq T_{\mathrm{eff}}^{(pq)}$.  No
thermodynamical meaning different from the one above is associated to the notion of effective temperatures here introduced. The global dynamics can be then readily interpreted in terms of well-known thermal-equilibrium physics by means of the various $T_{\mathrm{eff}}^{(nm)}$. However, out of thermal equilibrium, the various transitions feel in general different effective temperatures whose values depend on many factors such as the geometry of the body, the system-body distance and their interplay with the body optical resonances. By varying the various parameters one can tune the various effective temperatures far or close between them.

In the case $T_\textrm{W}=T_\textrm{M}=T$, in Eqs. \eqref{GammaN} and \eqref{n effective ij}, $n_{\mathrm{eff}}^{(nm)}=n(\omega_{nm},T)$ for any $(nm)$. It follows that the transition rates $\Gamma(\pm\omega_{nm})$ are factorized as a product of $\Gamma_0(\omega_{nm})$, of a term ($1+n(\omega_{nm},T)$ or $n(\omega_{nm},T)$) depending on temperature, times the factor $\alpha_\mathrm{W}(\omega_{nm})+\alpha_\mathrm{M}(\omega_{nm})$ depending on the properties of the body. Indeed, at thermal equilibrium the atomic decay rates depend on the presence of a material body because of the modification of the local field due to the field emitted by the body itself and to the way it scatters the one coming from the environment.

Equations \eqref{master equation N} and \eqref{GammaN}-\eqref{n effective ij} allow one to investigate the time evolution of the atomic density matrix in the presence of an arbitrary body and for any couple of temperatures $T_\text{M}$ and $T_\text{W}$. 

Remarkably, the explicit quantized form of $H_B$ of Eq. \eqref{Hamiltonian}  has not been needed here. Indeed, all quantities we are interested in are related to the fluctuations of the fields, which under the local equilibrium assumption are provided, as explained in this section, by the fluctuation-dissipation theorem derived only by using general properties of the fields.

In order to get an insight on the atomic dynamics we consider in the next two sections two specific examples in which the atom has a simple level structure.

\section{Two-level system}\label{par:application Two-level system}

\begin{figure}[h]
\scalebox{0.4}{\includegraphics{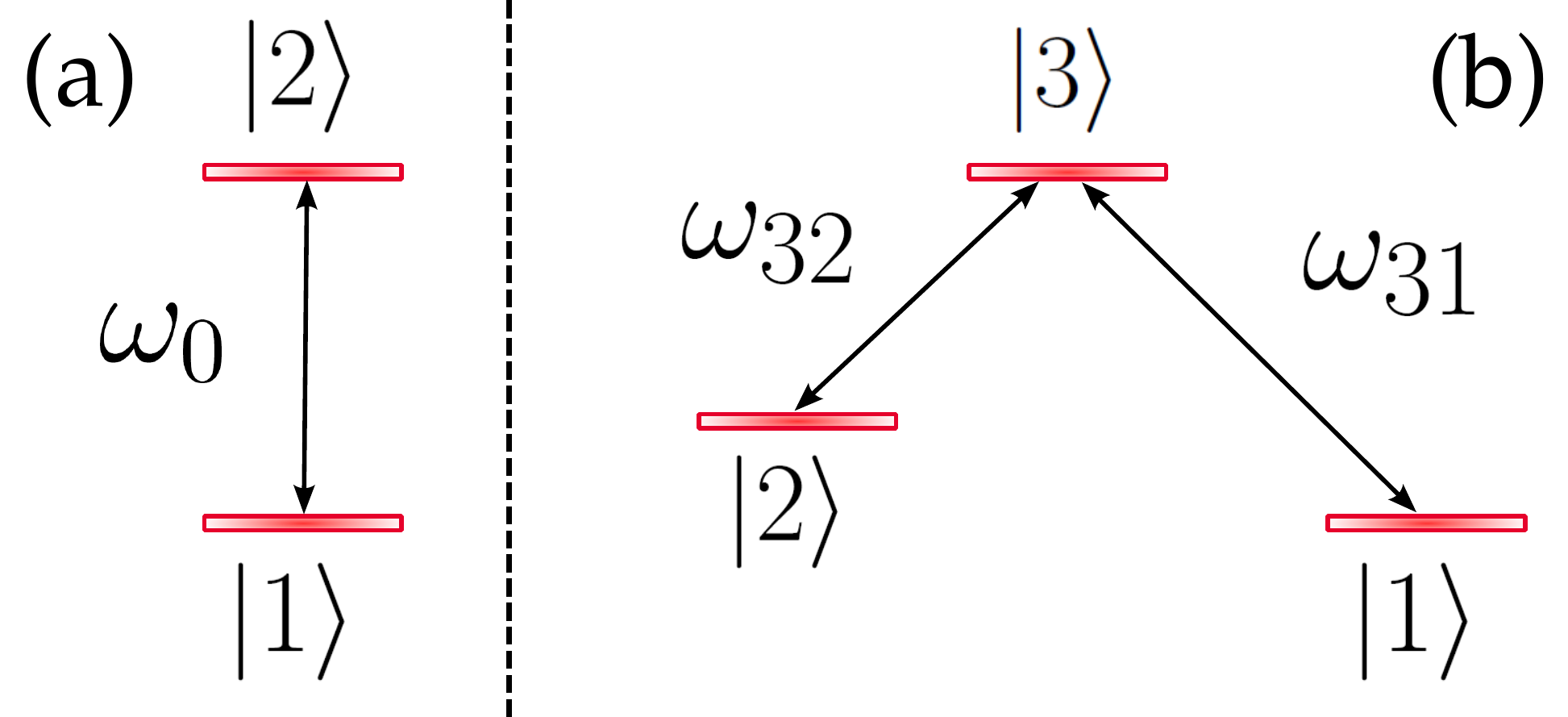}}
\caption{(color online). Scheme of levels and definition of transition frequencies for a two- (a) and three-level (b) atom.}
\label{FigSchemi}\end{figure}

In this section we assume that the atom has only two levels, $\omega_0=\omega_2-\omega_1$ being the transition frequency between the excited state $\ket{e}\equiv\ket{2}$ and the ground state $\ket{g}\equiv\ket{1}$ (see Fig. \ref{FigSchemi}(a)).

The free Hamiltonian of the two-level atom is
\begin{equation}\label{Atom hamiltonian}H_A=\sum_{n=1}^{2}{\hbar\omega_n}\ket{n}\bra{n},\end{equation}
and the atomic dipole operator $\mathbf{D}(t)$ assumes the simple form
\begin{equation}\label{dipole operator}\mathbf{D}(t)=\mathbf{d}\ket{1}\bra{2}e^{-i\omega_0 t}+\mathbf{d}^*\ket{2}\bra{1} e^{i\omega_0 t},\end{equation}
where $\mathbf{d}=\bra{g}\mathbf{D}\ket{e}$ (we assume that its diagonal matrix elements vanish). In Eqs. \eqref{Lamb shift Hamiltonian} and \eqref{dissipator term} the sum over $i$ and $j$ runs on $i,j =x, y, z$ and the sum over $\omega$ on the only two values $\omega_0$ and $-\omega_0$. By comparing Eqs. \eqref{Interaction hamiltonian}, \eqref{Interaction hamiltonian 2} and \eqref{dipole operator} one sees that $\mathbf{A}(\omega_0)=\mathbf{d}\ket{1}\bra{2}=\mathbf{A}^\dag(-\omega_0)$. By performing the sum over $\omega$, Eq. \eqref{master equation N} becomes
\begin{equation}\label{parameters master equation}\begin{split}&\frac{d}{d t}\rho(t)=-i\Bigl[\sum_n{\omega_n}\ket{n}\bra{n}+S(-\omega_0)\ket{1}\bra{1}\\
&\,+S(\omega_0)\ket{2}\bra{2},\rho(t)\Bigr]+\Gamma(-\omega_0)\Bigl(\rho_{11}\ket{2}\bra{2}-\frac{1}{2}\{\ket{1}\bra{1},\rho(t)\}\Bigr)\\
&\,+\Gamma(\omega_0)\Bigl(\rho_{22}\ket{1}\bra{1}-\frac{1}{2}\{\ket{2}\bra{2},\rho(t)\}\Bigr),\end{split}\end{equation}
where $S(-\omega_{0})$, $S(\omega_{0})$, $\Gamma(-\omega_{0})$, $\Gamma(\omega_{0})$ are defined as in Eq. \eqref{me parameters N}. The Lamb-shift Hamiltonian $H_{LS}=S(-\omega_0)\ket{1}\bra{1}+S(\omega_0)\ket{2}\bra{2}$ induces a shift in the eigenvalues of the free atom Hamiltonian $H_A$, which now become $\omega_1+S(-\omega_0)$ and $\omega_2+S(\omega_0)$ with a difference given by $\Delta_\omega=\omega_0+S(\omega_0)-S(-\omega_0)$. $\Gamma(\omega_0)$ and $\Gamma(-\omega_0)$ are the transition rates associated to the down- and upward transitions respectively. From Eq. \eqref{parameters master equation} the differential equations for $\rho_{ij}=\bra{i}\rho(t)\ket{j}$ follow:
\begin{equation}\begin{split}\frac{d}{dt}\rho_{11}(t)&=-\Gamma(-\omega_{0})\rho_{11}(t)+\Gamma(\omega_{0})\rho_{22}(t),\\
\frac{d}{dt}\rho_{22}(t)&=\Gamma(-\omega_{0})\rho_{11}(t)-\Gamma(\omega_{0})\rho_{22}(t)\\
\frac{d}{dt}\rho_{12}(t)&=\Bigl[i\Delta_{\omega}-\frac{\Gamma(\omega_{0})+\Gamma(-\omega_{0})}{2}\Bigr]\rho_{12}(t).\end{split}\end{equation}
The solution of this equation reads
\begin{equation}\label{rhoevolution}\begin{split}\begin{pmatrix}\rho_{11}(t)\\\rho_{22}(t)\end{pmatrix}&=\begin{pmatrix}\rho_{11}(0)\\\rho_{22}(0)\end{pmatrix}e^{-\gamma(\omega_0)t}+\frac{1-e^{-\gamma(\omega_0)t}}{\gamma(\omega_0)}\begin{pmatrix}\Gamma(\omega_0)\\\Gamma(-\omega_0)\end{pmatrix},\\
\rho_{12}(t)&=\rho_{12}(0)e^{i\Delta_\omega t}e^{-\frac{\gamma(\omega_0)}{2}t},\quad\rho_{21}(t)=\rho_{12}(t)^*.\end{split}\end{equation}
The decay rate $\gamma(\omega_0)=\Gamma(\omega_0)+\Gamma(-\omega_0)$ is not influenced by the Lamb shift, which does not play any role also in the steady state of the system. At times $t\gg1/\gamma(\omega_0)$ the system thermalizes to the steady state
\begin{equation}\label{stationary population excited state}\rho(t\to\infty)=\frac{1}{\Gamma(-\omega_0)+\Gamma(\omega_0)}\begin{pmatrix}\Gamma(\omega_0) & 0\\0 & \Gamma(-\omega_0)\end{pmatrix}.\end{equation}
The decoherence process is linked to the behavior of non-diagonal elements $\rho_{12}$ and $\rho_{21}$ and is essentially regulated by the decay rate $\gamma(\omega_0)/2$. As we will see, the behavior of the atomic evolution is qualitatively different between the equilibrium and non-equilibrium case.

\subsection{Case of thermal equilibrium}

When $T_\text{M}=T_\text{W}=T$, one has from Eq. \eqref{n effective ij} that $n_\text{eff}^{(21)}=n(\omega_0,T)$ so that the effective number of photons becomes independent of the material properties. Moreover, the above stationary state \eqref{stationary population excited state} reduces to
\begin{equation}\label{rhostattherm}\rho(t\to\infty)=\frac{1}{1+2n(\omega_0,T)}\begin{pmatrix}1+n(\omega_0,T) & 0\\0 & n(\omega_0,T)\end{pmatrix}.\end{equation}
As a consequence of the factorization of thermal and material dependence in $\Gamma(\pm\omega_0)$, there is a simplification of the factor $\alpha_\text{W}(\omega_0)+\alpha_\text{M}(\omega_0)$ in the steady state. It follows that the asymptotic state is a thermal state of a two-level system at temperature $T$, \emph{independently} of the presence and properties of body M. Out of thermal equilibrium, this peculiar cancellation does not hold anymore, the steady atomic state depending on the various parameters influencing  the behavior of the $\Gamma$ rates. We will show that the interplay between the body properties and the atomic position allows to realize a rich variety of asymptotic atomic configurations.

\subsection{Case out of thermal equilibrium}

When $T_\text{M}\neq T_\text{W}$, both the decay rates and the steady state of the system depend on the two temperatures and the properties of the body. In particular, these dependences are entangled in the term $n_\text{eff}^{(21)}$ of Eq. \eqref{n effective ij} (we pose in the two-level case $n_\text{eff}^{(21)}=n_\text{eff}$). Using Eq. \eqref{GammaN} in Eq. \eqref{stationary population excited state} we obtain
\begin{equation}\label{rhostat}\rho(t\to\infty)=\frac{1}{1+2n_\text{eff}}\begin{pmatrix}1+n_\text{eff} & 0\\0 & n_\text{eff}\end{pmatrix}.\end{equation}
We remark that the dependence on the material properties has not disappeared, being indeed contained in the effective number of photons $n_\text{eff}$. Nevertheless, the steady state of Eq. \eqref{rhostat} has the form of a thermal state so that the two-level system always thermalizes toward a thermal state even out of thermal equilibrium.
The temperature $T_\text{eff}$ of this steady state is defined as in Eq. \eqref{effective temperature ij} by $n(\omega_0,T_\text{eff})=n_\text{eff}$, it results equal to $k_{\rm B}T_\text{eff}=\hbar\omega/\ln(1+1/n_\text{eff})$ and is an intermediate temperature between $T_\text{M}$ and $T_\text{W}$.

We observe that the steady state \eqref{rhostat} is defined by only one parameter (i.e. one of the two diagonal elements) being $\Tr\rho=1$. As a consequence, any diagonal state satisfying the condition $\rho_{11}>\rho_{22}$ (always true in \eqref{rhostat}) coincides indeed with a thermal state. In the case of a system characterized by more than two levels, a diagonal state is defined by more than one parameter and thus can no longer be always identified with a thermal state, as we will see in the next section.

\section{Three-level system}\label{par:three level atom}

We now focus our attention on the case of a three-level atom in a $\Lambda$ configuration, previously considered in \cite{BellomoArxiv12}. We label the three states with $\ket{1}$, $\ket{2}$ and $\ket{3}$, with frequencies $\omega_1$, $\omega_2$ and $\omega_3$ (again in increasing order, see Fig. \ref{FigSchemi}(b)). We assume that the interaction Hamiltonian couples the states $\ket{1}$ and $\ket{3}$ as well as $\ket{2}$ and $\ket{3}$ ($\omega_{32}\neq \omega_{31}$), whereas the matrix element of $\mathbf{D}$ between $\ket{1}$ and $\ket{2}$ vanishes. The free Hamiltonian of the three-level atom is
\begin{equation}\label{Atom hamiltonian 3}H_A=\sum_{n=1}^3\hbar\omega_n\ket{n}\bra{n},\end{equation}
while the atomic dipole operator $\mathbf{D}(t)$ is given by
\begin{equation}\label{dipole operator 3}\begin{split}\mathbf{D}(t)&=\mathbf{d}_{13}\ket{1}\bra{3} e^{-i\omega_{31} t}+\mathbf{d}^*_{13}\ket{3}\bra{1} e^{i\omega_{31} t}\\
&\,+\mathbf{d}_{23}\ket{2}\bra{3} e^{-i\omega_{32} t}+\mathbf{d}^*_{23}\ket{3}\bra{2} e^{i\omega_{32} t},\end{split}\end{equation}
In Eqs. \eqref{Lamb shift Hamiltonian} and \eqref{dissipator term} the sum over $\omega$ runs now over the values $\omega_{31}$, $-\omega_{31}$, $\omega_{32}$ and $-\omega_{32}$. By performing the sum over $\omega_{nm}$ corresponding to the allowed transitions, Eq. \eqref{master equation N} reduces to
\begin{equation}\label{parameters master equation 3}\begin{split}&\frac{d}{dt}\rho=-i\Bigl[\sum_{n=1}^3{\omega_n}\ket{n}\bra{n}+S(-\omega_{31})\ket{1}\bra{1}\\
&\,+\Bigl(S(\omega_{31})+S(\omega_{32})\Bigr)\ket{3}\bra{3}+S(-\omega_{32})\ket{2}\bra{2},\rho\Bigr]\\
&\,+\Gamma(-\omega_{31})\Bigl(\rho_{11}\ket{3}\bra{3}-\frac{1}{2}\{\ket{1}\bra{1},\rho\}\Bigr)\\
&\,+\Gamma(\omega_{31})\Bigl(\rho_{33}\ket{1}\bra{1}-\frac{1}{2}\{\ket{3}\bra{3},\rho\}\Bigr)\\
&\,+\Gamma(-\omega_{32})\Bigl(\rho_{22}\ket{3}\bra{3}-\frac{1}{2}\{\ket{2}\bra{2},\rho\}\Bigr)\\
&\,+\Gamma(\omega_{32})\Bigl(\rho_{33}\ket{2}\bra{2}-\frac{1}{2}\{\ket{3}\bra{3},\rho\}\Bigr).\end{split}\end{equation}
The energy differences associated to the free Hamiltonian modified by the Lamb-shift term read $\Delta_{31}=\omega_{31}+S(\omega_{31})-S(-\omega_{31})+S(\omega_{32})$, $\Delta_{21}=\omega_{21}+S(-\omega_{32})-S(-\omega_{31})$ and $\Delta_{32}=\omega_{32}+S(\omega_{32})-S(-\omega_{32})+S(\omega_{31})$. The differential equations for the six diagonal and upper-diagonal matrix elements of the atomic density matrix are thus given by
\begin{equation}\begin{split}\frac{d}{dt}\rho_{11}(t)&=-\Gamma(-\omega_{31})\rho_{11}(t)+\Gamma(\omega_{31})\rho_{33}(t),\\
\frac{d}{dt}\rho_{22}(t)&=-\Gamma(-\omega_{32})\rho_{22}(t)+\Gamma(\omega_{32})\rho_{33}(t),\\
\frac{d}{dt}\rho_{33}(t)&=\Gamma(-\omega_{31})\rho_{11}(t)-\Gamma(\omega_{31})\rho_{33}(t)\\
&\,\:\:\:+\Gamma(-\omega_{32})\rho_{22}(t)-\Gamma(\omega_{32})\rho_{33}(t),\\
\frac{d}{dt}\rho_{12}(t)&=\Bigl[i\Delta_{21}-\frac{\Gamma(-\omega_{31})+\Gamma(-\omega_{32})}{2}\Bigr]\rho_{12}(t),\\
\frac{d}{dt}\rho_{13}(t)&=\Bigl[i\Delta_{31}-\frac{\Gamma(\omega_{31})+\Gamma(-\omega_{31})+\Gamma(\omega_{32})}{2}\Bigr]\rho_{13}(t),\\
\frac{d}{dt}\rho_{23}(t)&=\Bigl[i\Delta_{32}-\frac{\Gamma(\omega_{32})+\Gamma(-\omega_{32})+\Gamma(\omega_{31})}{2}\Bigr]\rho_{23}(t).\\\end{split}\end{equation}
The density matrix elements evolve asymptotically to
\begin{equation}\label{stationary population 3}\begin{split}&\rho_{11}(\infty)=\frac{\Gamma(\omega_{31})\Gamma(-\omega_{32})}{\tilde{Z}},\quad\rho_{22}(\infty)=\frac{\Gamma(-\omega_{31})\Gamma(\omega_{32})}{\tilde{Z}},\\
&\rho_{33}(\infty)=\frac{\Gamma(-\omega_{31})\Gamma(-\omega_{32})}{\tilde{Z}},\quad\rho_{ij}(\infty)=0\quad\text{for}\quad i\neq j,\\
&\tilde{Z}=\Gamma(\omega_{31})\Gamma(-\omega_{32})+\Gamma(-\omega_{31})\Gamma(\omega_{32})\\
&\quad+\Gamma(-\omega_{31})\Gamma(-\omega_{32}).\end{split}\end{equation}
As far as the decoherence process is concerned, it is linked to the behavior of nondiagonal elements $\rho_{12}$, $\rho_{13}$ and $\rho_{23}$, and then it is essentially regulated by the decay rates $\Gamma(\pm\omega_{31})$ and $\Gamma(\pm\omega_{32})$.

\subsection{Case of thermal equilibrium}

In the case $T_\textrm{W}=T_\textrm{M}=T$, using Eqs. \eqref{GammaN} and \eqref{n effective ij} and noticing that $n_{\mathrm{eff}}^{(31)}=n(\omega_{31},T)$ and $n_{\mathrm{eff}}^{(32)}=n(\omega_{32},T)$, the stationary state \eqref{stationary population 3} reduces to
\begin{equation}\label{stationary population thermal equilibrium}\begin{split}&\rho_{11}(\infty)=n(\omega_{32},T)\bigl[1+n(\omega_{31},T)\bigr]/Z_{\textrm{eq}}\\
&\rho_{22}(\infty)=n(\omega_{31},T)\bigl[1+n(\omega_{32},T)\bigr]/Z_{\textrm{eq}}\\
&\rho_{33}(\infty)=n(\omega_{31},T)n(\omega_{32},T)/Z_{\textrm{eq}}\\
&Z_{\textrm{eq}}=3 n(\omega_{31},T)n(\omega_{32},T)+n(\omega_{31},T)+n(\omega_{32},T).\\\end{split}\end{equation}
This state coincides indeed with the thermal state at temperature $T$ of a three-level system, as expected independentely of the properties of body M.

\subsection{Case out of thermal equilibrium}

The fundamental difference between the two-level and the three-level configuration is that in the latter we can associate an effective temperature comprised between $T_{\mathrm{W}}$ and $T_{\mathrm{M}}$ to each allowed transition, and that the two effective temperatures do not in general coincide. Equivalently, to each allowed transition one can associate an effective number of photons given by Eq. \eqref{n effective ij} for $(nm)\in\{(32),(31)\}$. The steady populations of Eq. \eqref{stationary population 3} can be rewritten in terms of the effective values of $n$ associated to the two transitions as follows:
\begin{equation}\label{steady state}\begin{split}&\begin{pmatrix}\rho_{11}(\infty)\\\rho_{22}(\infty)\\\rho_{33}(\infty)\end{pmatrix}=
\frac{1}{Z}\begin{pmatrix}n_{\mathrm{eff}}^{(32)}\bigl(1+n_{\mathrm{eff}}^{(31)}\bigr)\\\vspace{-.3cm}\\
n_{\mathrm{eff}}^{(31)}\bigl(1+n_{\mathrm{eff}}^{(32)}\bigr)\\\vspace{-.3cm}\\n_{\mathrm{eff}}^{(31)}n_{\mathrm{eff}}^{(32)}\end{pmatrix}\\
&\hspace{.5cm}Z=3n_{\mathrm{eff}}^{(31)}n_{\mathrm{eff}}^{(32)}+n_{\mathrm{eff}}^{(31)}+n_{\mathrm{eff}}^{(32)}.\end{split}\end{equation}
The decay rates $\Gamma(\pm\omega_{nm})$ have the form of Eq. \eqref{GammaN} for $(nm)\in\{(32),(31)\}$. Their values are always delimited by the equilibrium decay rates in the case $T_\mathrm{W}=T_\mathrm{M}=T_\mathrm{min} $ and in the case $T_\mathrm{W}=T_\mathrm{M}=T_\mathrm{max}$.

A first remark is that the steady state \eqref{steady state} is not in general a thermal state differently from the case of a two-level system. However a thermal state can be yet obtained if one succeeds in finding values of the various parameters such to make equal the two effective temperatures, $T_{\mathrm{eff}}^{(32)}=T_{\mathrm{eff}}^{(31)}=T$. This gives $n_{\mathrm{eff}}^{(31)}=n(\omega_{31},T)$ and $n_{\mathrm{eff}}^{(32)}=n(\omega_{32},T)$, so that the stationary state (\ref{steady state}) reduces to \eqref{stationary population thermal equilibrium}.

A second relevant remark is that while the quantities $\Gamma(\pm\omega_{nm})$, $n_{\mathrm{eff}}^{(nm)}$ and $T_{\mathrm{eff}}^{(nm)}$ associated to a given transition $(nm)$ are confined between their thermal-equilibrium values at $T_\mathrm{min}$ and $T_\mathrm{max}$, this is not the case for the steady populations {$\rho_{11}(\infty)$ and $\rho_{22}(\infty)$.} It is possible to show that when $n_{\mathrm{eff}}^{(32)}=n(\omega_{32}, T_\mathrm{max})$ and $n_{\mathrm{eff}}^{(31)}=n(\omega_{31},T_\mathrm{min})$,
$\rho_{11}(\infty)$ reaches its maximum value which results to be larger than its value when $T_\mathrm{W}=T_\mathrm{M}=T_\mathrm{min}$ (this is the maximum value for a thermal equilibrium configuration at a temperature comprised between $T_\mathrm{min}$ and $T_\mathrm{max}$). For this same condition $\rho_{22}(\infty)$ reaches its minimum value which is smaller than its value when $T_\mathrm{W}=T_\mathrm{M}=T_\mathrm{min}$. This peculiar behavior is connected to the fact that the condition $n_{\mathrm{eff}}^{(32)}=n(\omega_{32}, T_\mathrm{max})$ and $n_{\mathrm{eff}}^{(31)}=n(\omega_{31},T_\mathrm{min})$ makes at the same time the transition $\ket{2}\rightleftharpoons\ket{3}$ the most and $\ket{1}\rightleftharpoons\ket{3}$ the less reactive.
The opposite case is also of interest, namely when $T_{\mathrm{eff}}^{(\textrm{32})}=T_\mathrm{min}$ and $T_{\mathrm{eff}}^{(\textrm{31})}=T_\mathrm{max}$. In this case the minimum of $\rho_{11}(\infty)$ and the maximum of $\rho_{22}(\infty)$ are obtained. Once again this result is respectively smaller and larger than their equilibrium values at $T_\mathrm{W}=T_\mathrm{M}=T_\mathrm{max}$. In this last case $n_{\mathrm{eff}}^{(31)}=n(\omega_{31},T_\mathrm{max})$ and $n_{\mathrm{eff}}^{(32)}=n(\omega_{32},T_\mathrm{min})$, so that if $n(\omega_{32},T_\mathrm{min})<n(\omega_{31},T_\mathrm{max})$ there is population inversion. 

In the following we consider a specific example for which we give an explicit expression for $\alpha_\text{W}(\omega_{nm})$ and $\alpha_\text{M}(\omega_{nm})$ providing then a numerical investigation of the peculiar effects described above for an arbitrary body.

\section{Atom in front of a slab}\label{par: atom in front of a slab}

As a body we consider a slab of finite thickness $\delta$ for which explicit expressions for the transmission and reflection operators can be exploited \cite{MesAntEPL11,MesAntPRA11}. For this specific case, we will derive explicit analytic results for $\alpha_\mathrm{W}(\omega)$ and $\alpha_\mathrm{M}(\omega)$. The atom has position $\mathbf{R}=(0,0,z)$ (we have chosen $x=0$ and $y=0$ in virtue of the cylindrical symmetry of the problem with respect to the axis $z=0$) with $z>0$, whereas the slab is defined by the two interfaces $z=0$ and $z=-\delta$. As a result of the translational invariance of a planar slab with respect to the $xy$ plane, its reflection and transmission operators, $\mathcal{R}$ and $\mathcal{T}$, are diagonal and given by
\begin{equation}\label{RT1slab}\begin{split}\bra{p,\mathbf{k}}\mathcal{R}\ket{p',\mathbf{k}'}&=(2\pi)^2\delta(\mathbf{k}-\mathbf{k}')\delta_{pp'}\rho_{p}(\mathbf{k},\omega),\\
\bra{p,\mathbf{k}}\mathcal{T}\ket{p',\mathbf{k}'}&=(2\pi)^2\delta(\mathbf{k}-\mathbf{k}')\delta_{pp'}\tau_{p}(\mathbf{k},\omega).\end{split}\end{equation}
They are defined in terms of the Fresnel reflection and transmission coefficients modified by the finite thickness $\delta$ \cite{NotaTau}
\begin{equation}\begin{split}\rho_{p}(\mathbf{k},\omega)&=r_{p}(\mathbf{k},\omega)\frac{1-e^{2ik_{zm}\delta}}{1-r_{p}^2(\mathbf{k},\omega)e^{2ik_{zm}\delta}}\\
\tau_{p}(\mathbf{k},\omega)&=\frac{t_{p}(\mathbf{k},\omega)\bar{t}_{p}(\mathbf{k},\omega)e^{i(k_{zm}-k_z)\delta}}{1-r_{p}^2(\mathbf{k},\omega)e^{2ik_{zm}\delta}}.\\\end{split}\end{equation}
In these definitions we have introduced the $z$ component of the $\mathbf{K}$ vector inside the medium,
\begin{equation}k_{zm}=\sqrt{\varepsilon(\omega)\frac{\omega^2}{c^2}-\mathbf{k}^2},\end{equation}
$\varepsilon(\omega)$ being the dielectric permittivity of the slab, the ordinary vacuum-medium Fresnel reflection coefficients
\begin{equation}r_{\text{TE}}=\frac{k_z-k_{zm}}{k_z+k_{zm}},\qquad r_{\text{TM}}=\frac{\varepsilon(\omega)k_z-k_{zm}}{\varepsilon(\omega)k_z+k_{zm}},\end{equation}
as well as both the vacuum-medium (noted with $t$) and medium-vacuum (noted with $\bar{t}$) transmission coefficients
\begin{equation}\begin{split}t_{\text{TE}}&=\frac{2k_z}{k_z+k_{zm}},\qquad\hspace{.3cm}t_{\text{TM}}=\frac{2\sqrt{\varepsilon(\omega)}k_{z}}{\varepsilon(\omega)k_z+k_{zm}},\\
\bar{t}_{\text{TE}}&=\frac{2k_{zm}}{k_z+k_{zm}},\qquad\bar{t}_{\text{TM}}=\frac{2\sqrt{\varepsilon(\omega)}k_{zm}}{\varepsilon(\omega)k_z+k_{zm}}.\end{split}\end{equation}

After replacing the matrix elements \eqref{RT1slab} of the reflection and transmission operators in Eq. \eqref{alphas} we obtain, for an arbitrary transition frequency $\omega=\omega_{nm}$,
\begin{equation}\label{alphaslab}\begin{split}&\alpha_\text{W}(\omega)=\frac{3c}{8\pi\omega}\sum_{i,j}\frac{[\mathbf{d}_{mn}]_i^*[\mathbf{d}_{mn}]_j}{|\mathbf{d}_{mn}|^2}\sum_{p}\int_0^{2\pi}d\theta\Biggl\{\int_0^{\frac{\omega}{c}}\frac{dk\,k}{k_z}\\
&\,\times\Bigl[[\hat{\bbm[\epsilon]}_p^+(\mathbf{k},\omega)]_i[\hat{\bbm[\epsilon]}_{p}^{+}(\mathbf{k},\omega)]_j\bigl(|\rho_{p}(\mathbf{k},\omega)|^2+|\tau_{p}(\mathbf{k},\omega)|^2\bigr)\\
&\,+[\hat{\bbm[\epsilon]}_p^+(\mathbf{k},\omega)]_i[\hat{\bbm[\epsilon]}_{p}^{-}(\mathbf{k},\omega)]_je^{2ik_zz}\rho_{p}(\mathbf{k},\omega)\\
&\,+[\hat{\bbm[\epsilon]}_p^-(\mathbf{k},\omega)]_i[\hat{\bbm[\epsilon]}_{p}^{+}(\mathbf{k},\omega)]_je^{-2ik_zz}\rho_{p}(\mathbf{k},\omega)^*\\
&\,+[\hat{\bbm[\epsilon]}_p^-(\mathbf{k},\omega)]_i[\hat{\bbm[\epsilon]}_{p}^{-}(\mathbf{k},\omega)]_j\Bigr]\Biggr\},\\
&\alpha_\text{M}(\omega)=\frac{3c}{8\pi\omega}\sum_{i,j}\frac{[\mathbf{d}_{mn}]_i^*[\mathbf{d}_{mn}]_j}{|\mathbf{d}_{mn}|^2}\sum_{p}\int_0^{2\pi}d\theta\Biggl\{\int_0^{\frac{\omega}{c}}\frac{dk\,k}{k_z}\\
&\,\times[\hat{\bbm[\epsilon]}_p^+(\mathbf{k},\omega)]_i[\hat{\bbm[\epsilon]}_{p}^{+}(\mathbf{k},\omega)]_j\bigl(1-|\rho_{p}(\mathbf{k},\omega)|^2-|\tau_{p}(\mathbf{k},\omega)|^2\bigr)\\
&\,-i\int_{\frac{\omega}{c}}^{\infty}\frac{dk\,k}{\Ima(k_z)}[\hat{\bbm[\epsilon]}_p^+(\mathbf{k},\omega)]_i[\hat{\bbm[\epsilon]}_{p}^{+}(\mathbf{k},\omega)]_j^*\\
&\,\times e^{-2\Ima(k_z)z}\bigl(\rho_{p}(\mathbf{k},\omega)-\rho_{p}(\mathbf{k},\omega)^*\bigr)\Biggr\},\end{split}\end{equation}
where we have used that fact that the polarization vectors are real quantities in the propagative sector ($ck<\omega$). After performing explicitly the angular integrals, one obtains
\begin{equation}\label{alphaslab2}\begin{split}&\alpha_\text{W}(\omega)=\frac{3c}{8\omega}\sum_{i,j}\frac{[\mathbf{d}_{mn}]_i^*[\mathbf{d}_{mn}]_j}{|\mathbf{d}_{mn}|^2}\delta_{ij}\sum_p\Biggl\{\int_0^{\frac{\omega}{c}}\frac{k\,dk}{k_z}\\
&\,\times \Bigl[[\mathbf{M}_p^+(k,\omega)]_i\bigl(1+|\rho_{p}(k,\omega)|^2+|\tau_{p}(k,\omega)|^2\bigr)\\
&\,+2[\mathbf{M}_p^-(k,\omega)]_i\Rea\bigl(\rho_{p}(k,\omega)e^{2ik_zz}\bigr)\Bigr]\Biggr\},\\
&\alpha_\text{M}(\omega)=\frac{3c}{8\omega}\sum_{i,j}\frac{[\mathbf{d}_{mn}]_i^*[\mathbf{d}_{mn}]_j}{|\mathbf{d}_{mn}|^2}\delta_{ij}\sum_p\Biggl\{\int_0^{\frac{\omega}{c}}\frac{k\,dk}{k_z}\\
&\,\times[\mathbf{M}_p^+(k,\omega)]_i\bigl(1-|\rho_{p}(k,\omega)|^2-|\tau_{p}(k,\omega)|^2\bigr)\\
&\,+2\int_{\frac{\omega}{c}}^{+\infty}\frac{k\,dk}{\Ima(k_z)}e^{-2\Ima(k_z)z}[\mathbf{M}_p^+(k,\omega)]_i\Ima(\rho_{p}(k,\omega))\Biggr\},\end{split}\end{equation}
where
\begin{equation}\begin{split}\mathbf{M}_1^\phi(k,\omega)&=(1,1,0),\\
\mathbf{M}_2^\phi(k,\omega)&=\frac{c^2}{\omega^2}(\phi|k_z|^2,\phi|k_z|^2,2 k^2).\end{split}\end{equation}
In $\alpha_\text{W}(\omega_{nm})$ and $\alpha_\text{M}(\omega_{nm})$ only diagonal terms survive. They can be rewritten as
\begin{equation}\label{alphaE and alphaM}\begin{split}
\alpha_\mathrm{W}(\omega_{nm})&=\frac{\mathds{1}+\mathbf{B}(\omega_{nm})+2\mathbf{C}(\omega_{nm})}{2}\cdot\mathbf{\tilde{d}}_{nm},\\
\alpha_\mathrm{M}(\omega_{nm})&=\frac{\mathds{1}-\mathbf{B}(\omega_{nm})+2\mathbf{D}(\omega_{nm})}{2}\cdot\mathbf{\tilde{d}}_{nm},
\end{split}\end{equation}
where $\mathbf{\tilde{d}}_{nm}=(|[\mathbf{d}_{nm}]_x|^2,|[\mathbf{d}_{nm}]_y|^2,|[\mathbf{d}_{nm}]_z|^2)/|\mathbf{d}_{nm}|^2 $, $\mathds{1}=(1,1,1)$, we have used
\begin{equation}\begin{split}\label{integral A}
\sum_p\int_0^{\frac{\omega}{c}}\frac{k\,dk}{k_z}\mathbf{M}_p^+(k,\omega)=\frac{4\omega}{3 c}\mathds{1},\end{split}\end{equation}
and introduced the vectors
\begin{equation}\begin{split}\label{integrals}
\mathbf{B}(\omega)&=\frac{3c}{4\omega}\sum_p\int_0^{\frac{\omega}{c}}\frac{k\,dk}{k_z}\mathbf{M}_p^+(k,\omega)\\
&\,\times\bigl(|\rho_{p}(k,\omega)|^2+|\tau_{p}(k,\omega)|^2\bigr),\\
\mathbf{C}(\omega)&=\frac{3c}{4\omega}\sum_p\int_0^{\frac{\omega}{c}}\frac{k\,dk}{k_z}\mathbf{M}_p^-(k,\omega)\Rea\bigl(\rho_{p}(k,\omega)e^{2ik_zz}\bigr),\\
\mathbf{D}(\omega)&=\frac{3c}{4\omega}\sum_p\int_{\frac{\omega}{c}}^{+\infty}\!\!\!\!\frac{k\,dk}{\Ima(k_z)}e^{-2\Ima(k_z)z}\mathbf{M}_p^+(k,\omega)\\
&\,\times\Ima(\rho_{p}(k,\omega)).\\\end{split}\end{equation}
The integrals appearing in $\mathbf{B}$ and $\mathbf{C}$ are defined in the propagative sector, while $\mathbf{D}$ is the only evanescent contribution. $\mathbf{B}$ is distance-independent and depends on the thickness $\delta$, while $\mathbf{C}$ and $\mathbf{D}$ depend both on $\delta$ and on the atom-slab distance $z$. Moreover, it is possible to show that $\mathbf{C}$ and $\mathbf{D}$ tend to 0 for $z\to\infty$, while for $z\to0$ $\mathbf{C}$ remains finite and $\mathbf{D}$ diverges (more precisely, its TM contribution) as (see Appendix \ref{par:integral} where the dependence of $\mathbf{B}(\omega)$, $\mathbf{C}(\omega)$ and $\mathbf{D}(\omega)$ on $\delta$ and $z$ as well as the asymptotic behavior of $\mathbf{D}(\omega)$ are discussed)
\begin{equation}\mathbf{D}(\omega,z\to0)\simeq\frac{3c^3}{16\omega^3z^3}\Ima\left(\frac{\varepsilon(\omega)-1}{\varepsilon(\omega)+1}\right)(1,1,2).\end{equation}
It follows from Eq. \eqref{alphaE and alphaM} that for small $z$ $\alpha_\text{M}(\omega_{nm})\gg\alpha_\text{W}(\omega_{nm})$ because of the divergence of $\mathbf{D}$. As a consequence, all the effective temperatures tend to $T_\text{M}$ in this limit (see Eqs. \eqref{n effective ij} and \eqref{effective temperature ij}) and thus the atom thermalizes at the body temperature $T_\text{M}$. On the contrary, for large $z$, $\alpha_\text{W}(\omega_{nm})>\alpha_\text{M}(\omega_{nm})$ and the analysis of Eqs. \eqref{n effective ij} and \eqref{effective temperature ij} shows that the position of each $T_\text{eff}^{(nm)}$ in the interval $[T_\text{min},T_\text{max}]$ is governed by $\mathbf{B}(\omega_{nm})$. From these comments it follows that it always exists a distance $z$ for which $\alpha_\text{W}(\omega_{nm})=\alpha_\text{M}(\omega_{nm})$. At this position $n_\text{eff}^{(nm)}$ does not depend on the geometry of the system and of the dielectric properties of the body (even if the point where this happens 
depends on them). This point delimits the two zones of influence where each temperature dominates for that specific transition. We also note that at thermal equilibrium $\mathbf{B}$ does not play any role, since all the quantities are proportional to the sum $\alpha_\text{W}+\alpha_\text{M}$, independent of $\mathbf{B}$ and asymptotically equal to 1. We remark that since $\Gamma(-\omega_{nm})$ (see Eq. \eqref{GammaN}) must be always larger than zero, by considering the cases $T_\textrm{W}=T_\textrm{M}$, $T_\textrm{W}=0, T_\textrm{M}\neq 0$ and $T_\textrm{W}\neq 0, T_\textrm{M}=0$ three constraints follow (only two of them are independent), for $i=x,y,z$:
\begin{equation}\label{constrains}\begin{split}1+[\mathbf{B}(\omega)]_i+2[\mathbf{C}(\omega)]_i&\ge0,\\
1-[\mathbf{B}(\omega)]_i+2[\mathbf{D}(\omega)]_i&\ge0,\\
1+[\mathbf{C}(\omega)]_i+[\mathbf{D}(\omega)]_i&\ge 0.\end{split}\end{equation}

We note that the three relevant special cases in which the electric dipole is parallel ($|[\mathbf{d}_{nm}]_x|^2=|[\mathbf{d}_{nm}]_y|^2=|[\mathbf{d}_{nm}]|^2/2,[\mathbf{d}_{nm}]_z=0$) or perpendicular ($[\mathbf{d}_{nm}]_x=[\mathbf{d}_{nm}]_y=0, |[\mathbf{d}_{nm}]_z|^2=|[\mathbf{d}_{nm}]|^2$) to the surface, or isotropic ($|[\mathbf{d}_{nm}]_i|^2=|[\mathbf{d}_{nm}]|^2/3$ for $i=x,y,z$), can be easily treated by specifying the symmetries of electric-dipole matrix elements in Eq. \eqref{alphas} or in Eq. \eqref{alphaE and alphaM}.

Before providing a numerical investigation for both cases of two- and three-level atom, we briefly discuss some limiting cases already treated in literature. The case in which the body is absent is obtained by putting, in the general discussion of Sec. \ref{par:one body}, $\mathcal{T}=1$, $\mathcal{R} =0$, or equivalently in Eq. \eqref{integrals} $\rho_{p}(\mathbf{k},\omega)=0$ and $\tau_{p}(\mathbf{k},\omega)=1$ ($\varepsilon(\omega)=1$), leading to $\mathbf{B}(\omega)=\mathds{1}$ and $\mathbf{C}(\omega)=\mathbf{D}(\omega)=\mathbf{0}$. It follows that $\alpha_\text{W}(\omega)=1$ and $\alpha_\text{M}(\omega)=0$, so that the rates $\Gamma(\pm\omega_{nm})$ of Eq. \eqref{GammaN} do not depend on the properties of the atomic dipole, and are equal to $\Gamma(\omega)=\Gamma_0\left[1+n(\omega,T_\textrm{W})\right]$ and $\Gamma(-\omega)=\Gamma_0n(\omega,T_\textrm{W})$. As expected, the body temperature $T_\text{M}$ disappears.

In the limiting case of thermal equilibrium in the presence of the slab, when $T_\textrm{W}=T_\textrm{M}=T$, Eq. \eqref{GammaN} can be recast as
\begin{equation}\label{GammaNslabeq}\begin{split}\begin{pmatrix}\Gamma(\omega_{nm})\\\Gamma(-\omega_{nm})\end{pmatrix}&=\Gamma_0(\omega_{nm})\Bigl[\mathds{1}+\mathbf{C}(\omega_{nm})+\mathbf{D}(\omega_{nm})\Bigr]\cdot\mathbf{\tilde{d}}_{nm}\\
&\,\times\begin{pmatrix}1+n(\omega,T)\\n(\omega,T)\end{pmatrix}.\end{split}\end{equation}
We notice that in the limit of zero temperature Eq. \eqref{GammaNslabeq} gives the known-result for the modification of the atomic decay rate in front of a slab reported in \cite{Novotny06,BiehsPRA11}.

When the slab is a perfect mirror, one has $\varepsilon(\omega)\rightarrow\infty$, $\rho_{p}(\mathbf{k},\omega)=(-1)^p$ and $\tau_{p}(\mathbf{k},\omega)=0$. In this case the temperature of the slab does not play any role. This corresponds to the fact that under this assumption (dielectric permittivity real and infinite over the entire frequency spectrum) a temperature cannot be properly defined. By using in Eq. \eqref{integrals} $\rho_{p}(\mathbf{k},\omega)=(-1)^p$ and $\tau_{p}(\mathbf{k},\omega)=0$, we have $\mathbf{B}(\omega)=\mathds{1}$, $\mathbf{D}(\omega)=\mathbf{0}$ and
\begin{equation}\label{Cmirror}\begin{split}[\mathbf{C}(\omega)]_i&=\frac{3c}{4\omega}\sum_p(-1)^p\int_0^{\frac{\omega}{c}}\frac{k\,dk}{k_z}[\mathbf{M}_p^-(k,\omega)]_i\cos\bigl(2k_zz\bigr)\\
&=\frac{3}{2}\Bigl[\Bigl(\frac{\sin\tilde{z}}{\tilde{z}^3}-\frac{\cos\tilde{z}}{\tilde{z}^2}\Bigr)\bigl(\delta _{ix}+\delta_{iy}+2\delta_{iz}\bigr)\\
&\,-\frac{\sin\tilde{z}}{\tilde{z}}\bigl(\delta _{ix}+\delta_{iy}\bigr)\Bigr]\end{split}\end{equation}
where $\tilde{z}=\frac{2z\omega}{c}$. Equation \eqref{GammaN} reduces to
\begin{equation}\label{GammaNmirror}\begin{split}\begin{pmatrix}\Gamma(\omega_{nm})\\\Gamma(-\omega_{nm})\end{pmatrix}&=\Gamma_0(\omega_{nm})\Bigl[\mathds{1}+\mathbf{C}(\omega_{nm})\Bigr]\cdot \mathbf{\tilde{d}}_{nm}\\
&\,\times\begin{pmatrix}1+n(\omega_{nm},T_\textrm{W})\\n(\omega_{nm},T_\textrm{W})\end{pmatrix}.\end{split}\end{equation}
Equations \eqref{Cmirror} and \eqref{GammaNmirror} allow one to retrieve known results for this ideal case \cite{Novotny06}. For example, for a dipole moment parallel to the surface the emission rate is reduced to zero close to the slab, while in the case the dipole moment is perpendicular to the surface it is enhanced with a factor 2 with respect to the absence of the slab.

\section{Numerical analysis}\label{par:Numerical analysis}

In this section we provide numerical analysis for both cases of two- and three-level atoms treated respectively in Secs. \ref{par:application Two-level system} and \ref{par:three level atom}, considering an atom in front of a slab made of gallium arsenide (GaAs). The dielectric permittivity $\epsilon (\omega)$ of GaAs is described using a Drude-Lorentz model \cite{Palik98}
\begin{equation}\epsilon (\omega) =\epsilon_{\textrm{inf}}\frac{\omega ^2-\omega_l^2+i\Gamma\omega}{\omega^2-\omega_r^2+i\Gamma\omega},\end{equation}
characterized by a resonance at $\omega_r=0.506\times10^{14}\,\mathrm{rad}\,\text{s}^{-1}$ and where $\epsilon_{\textrm{inf}}=11 $, $\omega_l=0.550\times10^{14}\,\mathrm{rad}\,\text{s}^{-1}$ and $\Gamma=0.00452\times10^{14}\,\mathrm{rad}\,\text{s}^{-1}$. This model implies a surface phonon-polariton resonance at $\omega_p=0.547\times10^{14}\,\mathrm{rad}\,\text{s}^{-1}$. A relevant length scale in this case is $c/\omega_r\simeq 6\,\mu$m while a reference temperature is $\hbar\omega_r/k_B\simeq 387\,$K. The main advantage of GaAs with respect to silicon carbide (SiC), previously used for the numerical analysis in \cite{BellomoArxiv12}, is that its optically relevant frequencies are smaller than those of SiC. This allows one to observe interesting effects in a domain of frequencies where the thermal populations $n(\omega,T)$ are larger. Furthermore, in order to discuss the dependence of atomic dynamics with respect to changes of the overall optical response of the slab, we also briefly consider the case of a
metallic slab. We specify here that all the numerical calculations presented in this section refer to the case of isotropic dipoles.

As we will see, the atomic frequencies will be chosen of the same order of the slab resonances. In the case of a GaAs slab this can be achieved, for example, by considering as an atomic system self-assembled quantum dots made of InGaAs, characterized by transition frequencies in the few terahertz
range  \cite{Zibik09}. In these systems, the transition frequency can be modified by varying the size and the composition of the quantum dot.  More in general, given an atomic system, one can look at materials for the body whose resonances match the atomic frequencies.

\subsection{Transition rates and effective temperatures}\label{par:common}

We first provide the analysis of quantities which are indeed common to the case of a two- and a three-level atom, namely the transition rates $\Gamma(\pm\omega)$ and the effective temperature for an arbitrary frequency $\omega$.

\begin{figure}[h]
\scalebox{0.59}{\includegraphics{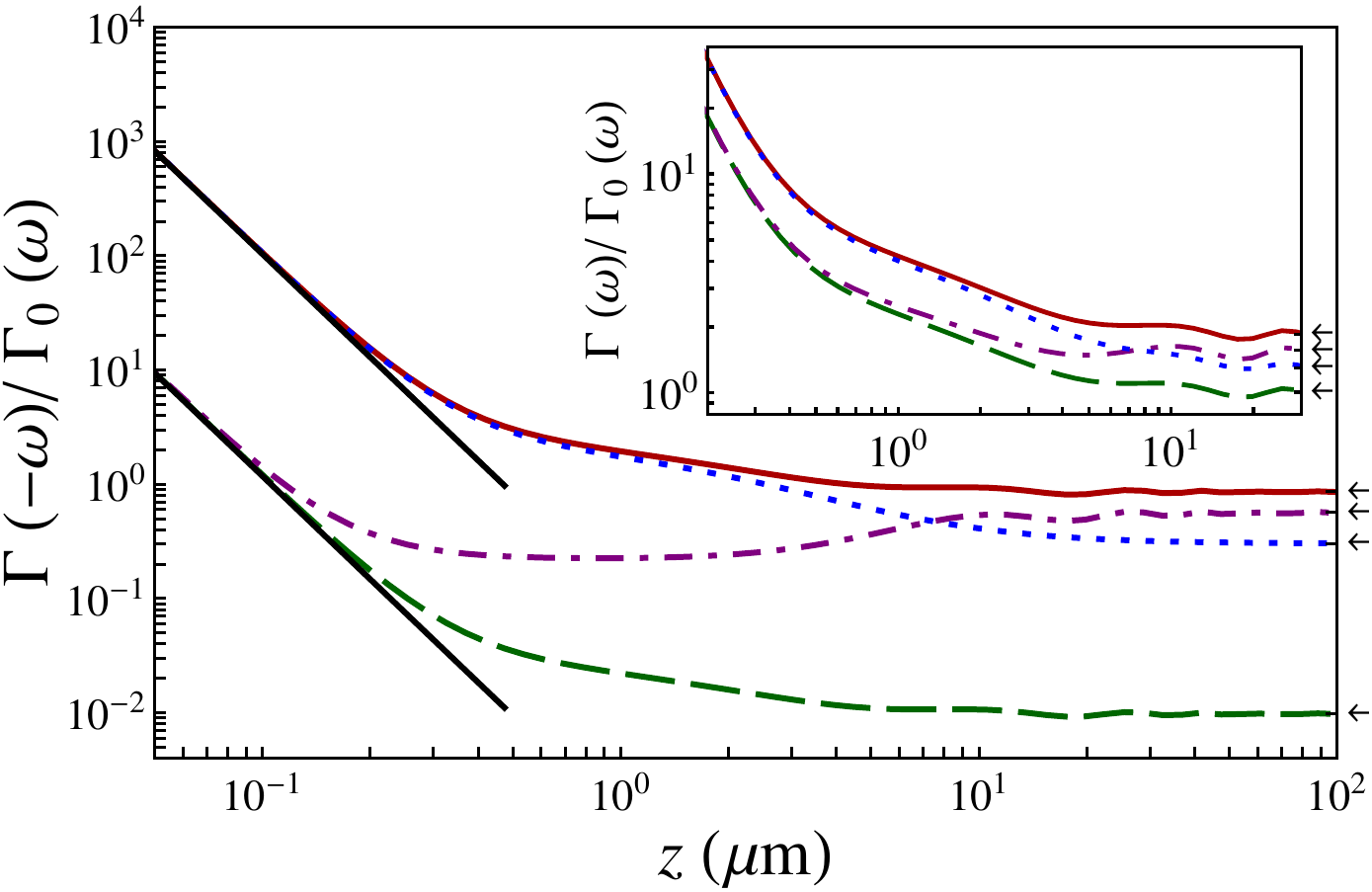}}
\caption{(Color online) Main part: $\Gamma(-\omega)/\Gamma_0(\omega)$ as a function of $z$ for $\omega=1.2\omega_r$ and $\delta=1$ cm (semi-infinite slab)  for $(T_\text{min},T_\text{max})=(100, 600)$\,K. Asymptotic curves for small $z$ (black segments) for $T_\text{M}=T_\text{W}=T_\text{max}$ and $T_\text{M}=T_\text{W}=T_\text{min}$. Inset: $\Gamma(\omega)/\Gamma_0(\omega)$ as a function of $z$. For both the main part and the inset: $T_\text{M}=T_\text{W}=T_\text{max}$ (red solid line), $(T_\text{W},T_\text{M})=(T_\text{max},T_\text{min})$ (purple dot-dashed line), $(T_\text{W},T_\text{M})=(T_\text{min},T_\text{max})$ (blue dotted line), $T_\text{M}=T_\text{W}=T_\text{min}$ (green dashed line). The symbols $\leftarrow$ indicate the asymptotic values (with respect to $z$) corresponding to the four couples of temperatures (see Appendix \ref{par:integral}).}
\label{Fig2}\end{figure}

In Fig. \ref{Fig2} we plot $\Gamma(-\omega)/\Gamma_0(\omega)$ and $\Gamma(\omega)/\Gamma_0(\omega)$ as a function of the atom-slab distance $z$. For a given couple of temperatures $(T_\text{min},T_\text{max})=(100, 600)$\,K, we compare the two thermal-equilibrium configurations at $100\,$K and $600\,$K with the two possible configurations out of thermal-equilibrium. From the plot we see that out of thermal equilibrium the transition rates are always confined between the corresponding thermal-equilibrium values. For small values of $z$ only the temperature of the slab contributes. On the contrary, we see that the asymptotic value reached for large values of $z$ differs from the equilibrium realization at the environmental temperature. This means that even at large atom-slab separations both temperatures play a role in the atomic dynamics.

\begin{figure}
\scalebox{0.54}{\includegraphics{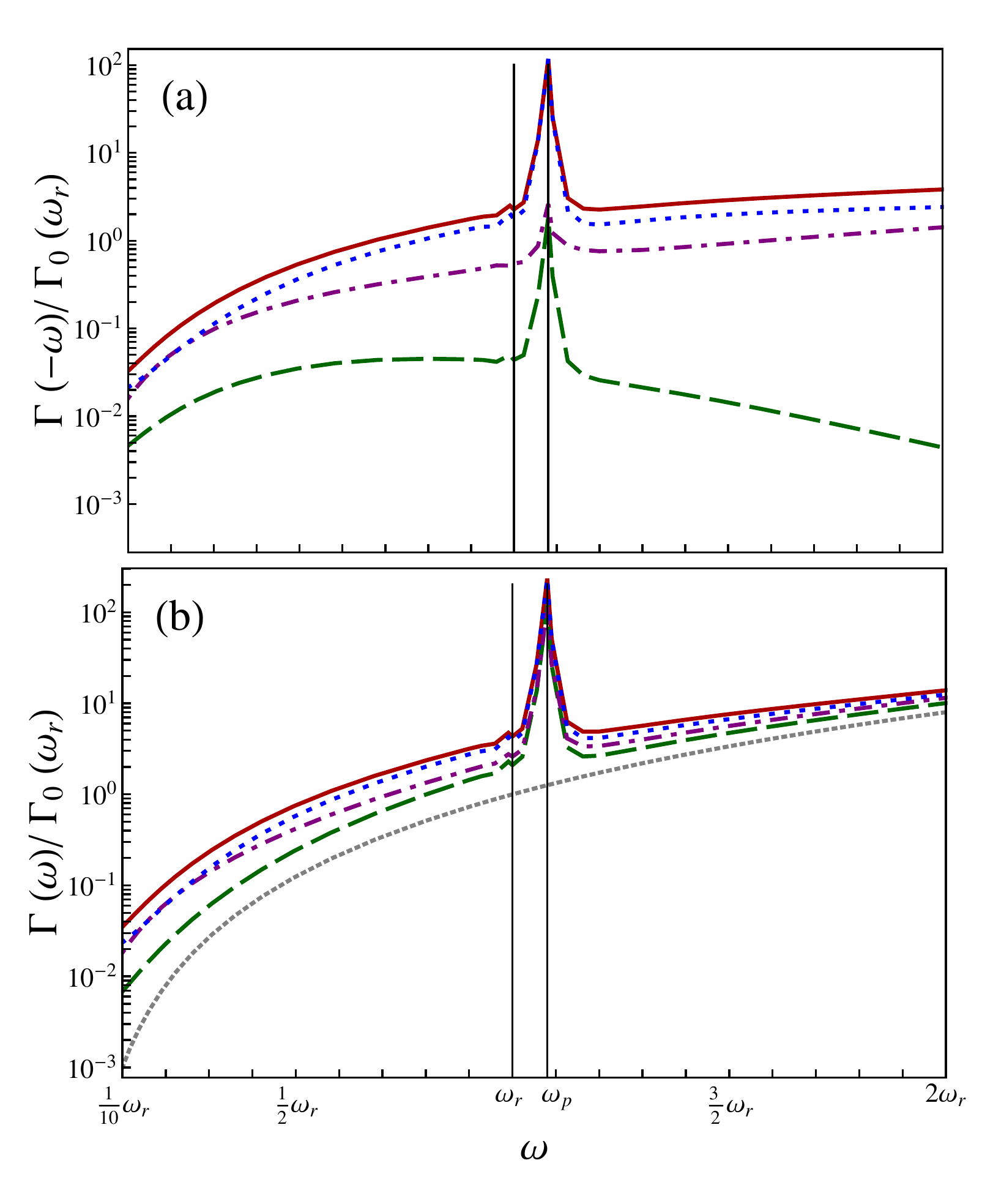}}
\caption{(color online). Panel (a): $\Gamma(-\omega)/\Gamma_0(\omega_r)$ as a function of $\omega$. Panel (b): $\Gamma(\omega)/\Gamma_0(\omega_r)$ and $\Gamma_0(\omega)/\Gamma_0(\omega_r)$ (gray small dotted line) as a function of $\omega$. Same color convention of Fig. \ref{Fig2} for $(T_\text{min},T_\text{max})=(100, 600)$\,K. For both panels $z=1\,\mu$m and $\delta=1.4\,\mu$m.}
\label{Fig2b}\end{figure}

Let us now discuss the dependence on the atomic transition frequency $\omega$ for $(T_\text{min},T_\text{max})=(100, 600)$\,K. Figure \ref{Fig2b} shows that, both at and out of thermal equilibrium the rates $\Gamma$ have a particularly pronounced dependence on $\omega$ around $\omega_p$.
In this sense the atomic dynamics changes strongly if the transition frequency is tuned around the frequencies which are relevant for the dielectric response of the body ($\omega_p$ in particular). Figure \ref{Fig2b} also shows that {$\Gamma(-\omega)$} allows a much wider range of variations between the two thermal-equilibrium cases. This is due to the fact that, while $\Gamma(-\omega)$ is directly proportional to $n_\text{eff}$, the dependence of $\Gamma(\omega)$ on $1+n_\text{eff}$ greatly reduces the relative changes with respect to $T_\text{M}$ and $T_\text{W}$.

\begin{figure}[h]
\includegraphics[width=0.48\textwidth]{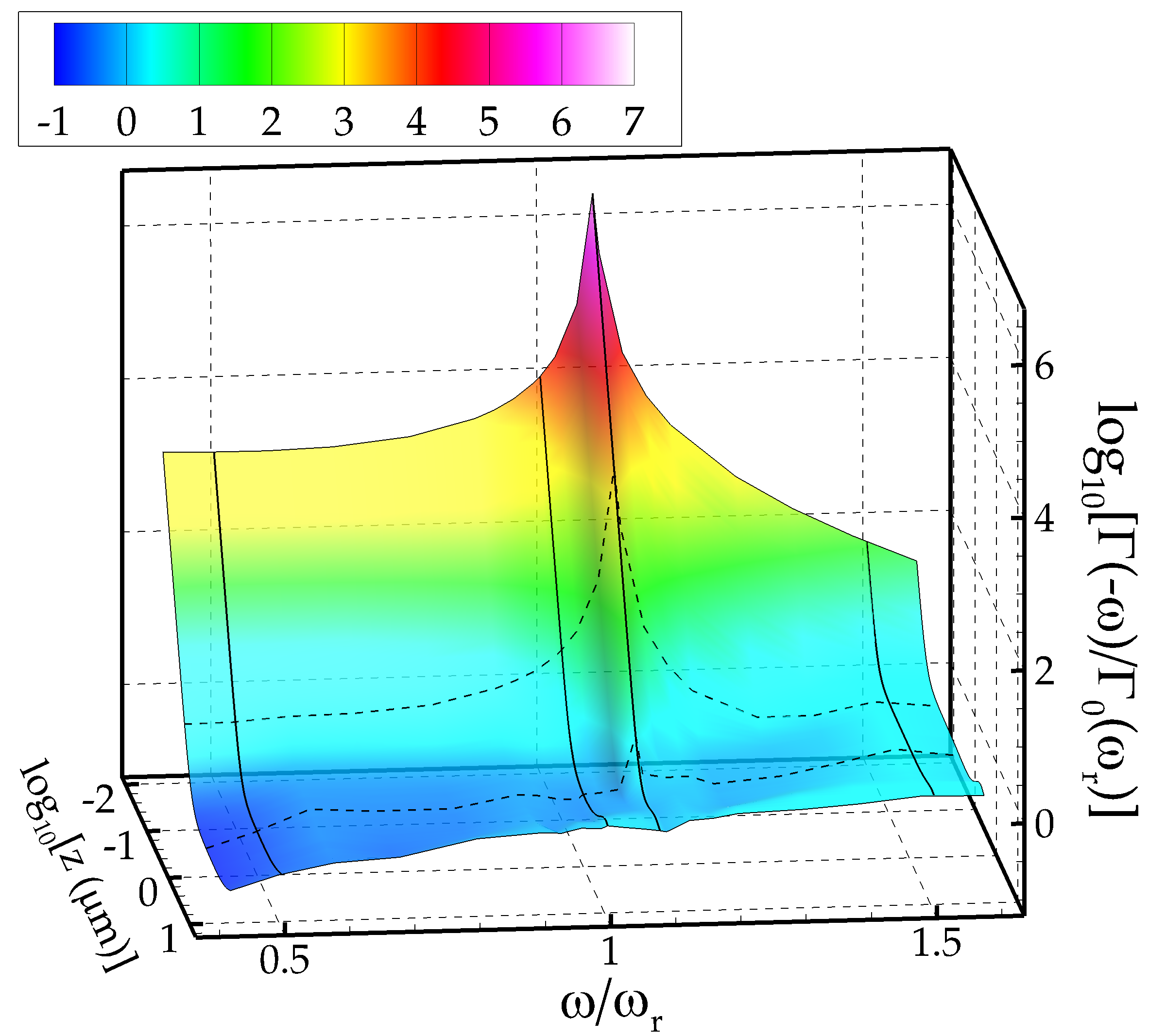}
\caption{(color online). $\Gamma(-\omega)/\Gamma_0(\omega)$ as a function of $z$ and $\omega$ for $\delta=8.4\,\mu$m and $(T_\text{W},T_\text{M})=(600,100)\,$K.}
\label{Fig3}\end{figure}

In Fig. \ref{Fig3} we plot $\Gamma(-\omega)/\Gamma_0(\omega_r)$ as a function of $z$ and $\omega$ for $\delta=8.4\,\mu$m and $(T_\text{W},T_\text{M})=(600,100)\,$K. The plot elucidates the role of slab resonances $\omega_r$ and $\omega_p$ in the behavior of the transition rates. In particular, we observe that for any value of the atom-slab distance $z$, the transition rates always show a peak centered at $\omega_p$.

\begin{figure}[h]
\scalebox{0.052}{\includegraphics{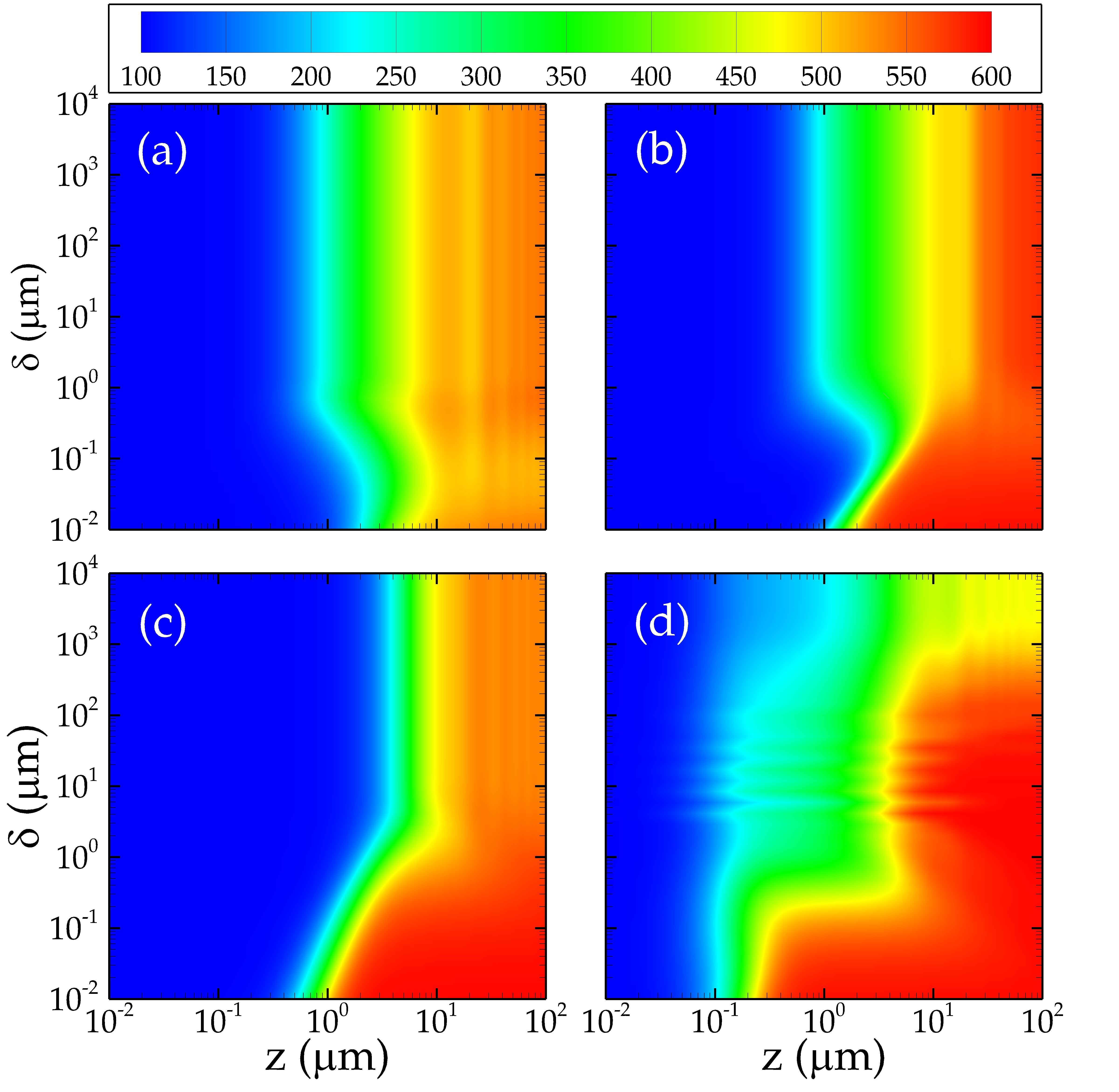}}
\caption{(color online). Density plot of $T_{\mathrm{eff}}^{(nm)}$ as a function of $z$ and $\delta$ for four different frequencies for $(T_\text{W},T_\text{M})=(600,100)\,$K. The chosen frequencies are: $\omega_r$ (a), $1.02\,\omega_r$ (b), $\omega_p$ (c), and $1.5\,\omega_r$ (d).}
\label{Fig4}\end{figure}

In Fig. \ref{Fig4} the effective temperature $T_{\mathrm{eff}}^{(nm)}$ as a function of $z$ and $\delta$ is plotted for four different frequencies. The temperatures chosen are again $(T_\text{W},T_\text{M})=(600,100)\,$K. The plot shows that by varying the atomic transition frequency $\omega_{nm}$ the effective temperatures may behave quite differently, with the common feature that for $z$ small enough only the slab temperature remains relevant. For large $z$ both temperatures contribute in a way which depends both on the frequency and the slab thickness $\delta$. For intermediate distances the remaining range of temperatures is explored. We remark that small variations of frequency around the slab resonances (compare for example \ref{Fig4}(a) and \ref{Fig4}(b)) may lead to substantial changes in the behavior of $T_{\mathrm{eff}}^{(nm)}$. The comparison between the various plots may allow one to identify regions where the various transitions feel temperatures far between each other.

\subsection{Two-level atom}\label{par:NAqubit}

\begin{figure}[h]
\includegraphics[width=0.50\textwidth]{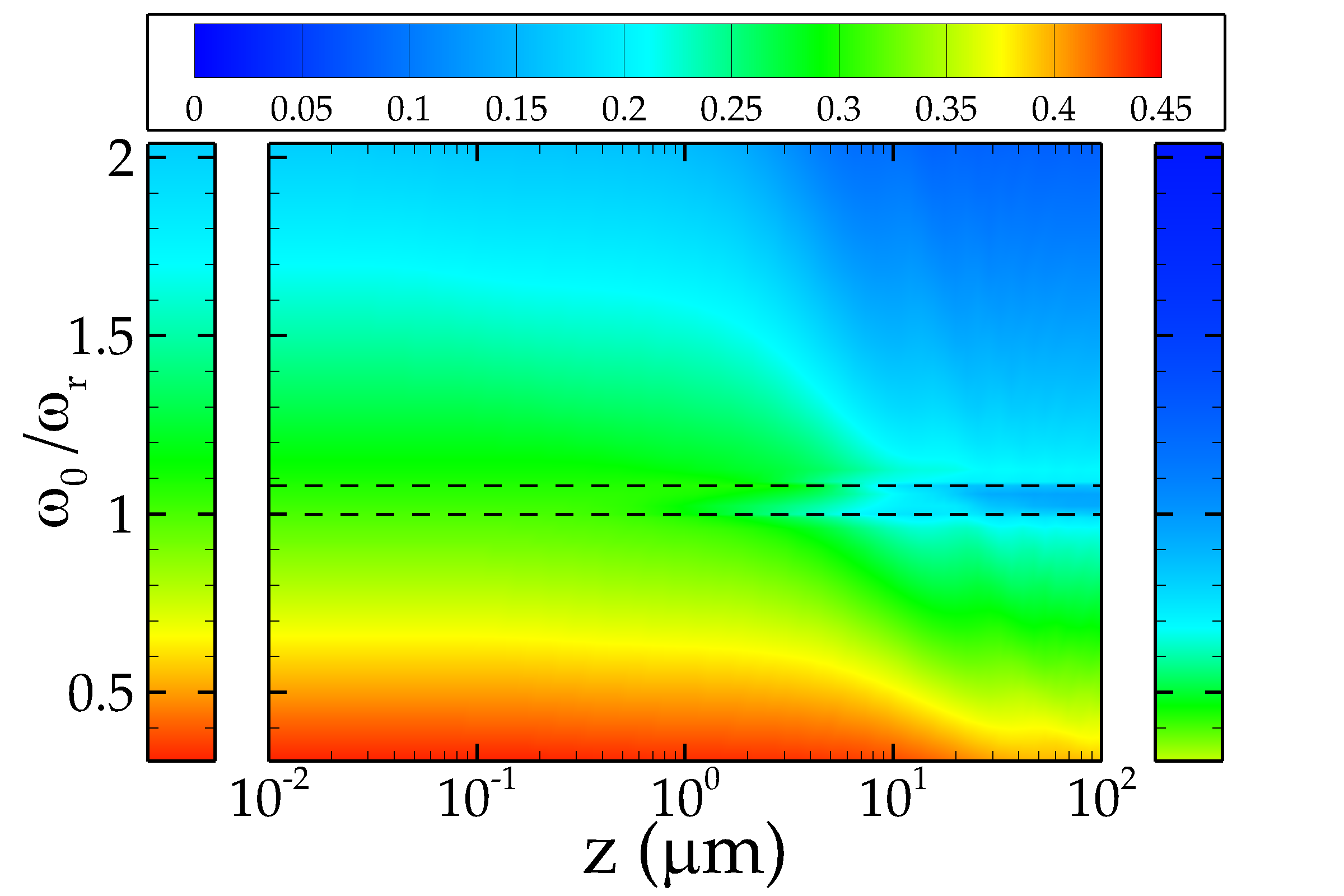}
\caption{(color online). Density plot of the excited-state population $\rho_{22} (\infty)$ as a function of $z$ and $\omega_0$ for $\delta=1\,$cm, corresponding to a semi-infinite slab. The temperatures are $(T_\text{W},T_\text{M})=(200,500)\,$K. On the left (right) of the main part, the density plot at thermal equilibrium at 500\,K (200\,K) is given.}
\label{Fig5}\end{figure}

We now focus our attention on the case of a two-level atom. For this configuration, the steady state of the system is of course identified by only one parameter, for example the excited-state population $\rho_{22}(\infty)$. In Fig. \ref{Fig5} we plot this population as a function of $z$ and $\omega_0$ for $\delta=1\,$cm (semi-infinite slab). In the main part, the chosen temperatures are $(T_\text{W},T_\text{M})=(200,500)\,$K. On the left (right) of the main part, the density plot at thermal equilibrium at 500\,K (200\,K) is given, which is independent of $z$. The main part  shows that out of thermal equilibrium $\rho_{22}$ depends on the atomic position. From the comparison of the three plots, we confirm that for small $z$ only the slab temperature contributes, while for large $z$ both $T_\text{W}$ and $T_\text{M}$ are relevant to determine the values of the population. The dashed lines (corresponding to $\omega_r$ and $\omega_p$) highlight that the variations of the atomic population are paricularly
pronounced around the slab resonances.

\begin{figure}[h]
\includegraphics[width=0.48\textwidth]{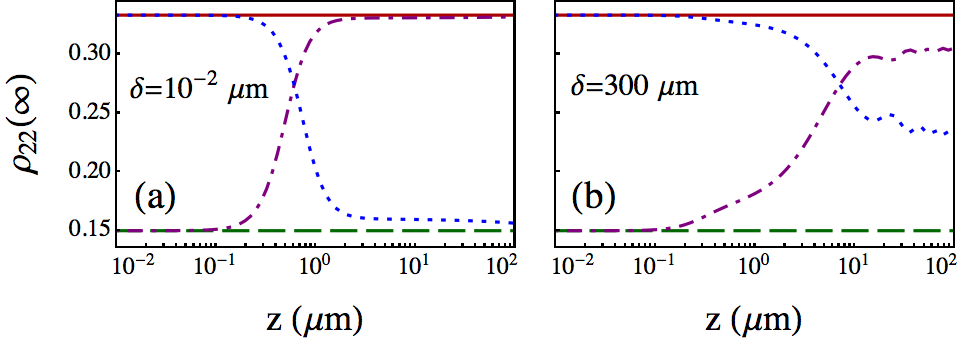}
\caption{(color online). $\rho_{22}(\infty)$ as a function of $z$ for $\omega_0=0.9\,\omega_r$, $(T_\text{min},T_\text{max})=(200, 500)$\,K. Same color convention of Fig. \ref{Fig2}. The two considered thicknesses are $\delta=0.01\,\mu$m (a) and $\delta=300\,\mu$m (b).}
\label{Fig5a}\end{figure}

In Fig. \ref{Fig5a} $\rho_{22}(\infty)$ is plotted as a function of $z$ for two values of $\delta$ at $\omega_0=0.9\,\omega_r$ 
for $(T_\text{min},T_\text{max})=(200, 500)$\,K.
We see that for $\delta=0.01\,\mu$m the atom thermalizes at large distance $z$ at a temperature very close to $T_\text{M}$ (see Fig. \ref{Fig5a}(a)), while by increasing the thickness to $300\,\mu$m (limit of semi-infinite slab) one obtains a steady state at an intermediate temperature between $T_\text{M}$ and $T_\text{W}$ (see Fig. \ref{Fig5a}(b)). This shows that the slab thickness $\delta$ plays a role in defining the region of influence of the slab temperature.

The mathematical reason for which in close proximity of the body the atom tends to thermalize to the temperature of the body is the divergence of the integral  $ \mathbf{D}(\omega)$ of Eq. \eqref{integrals}, corresponding to the fact that the field is dominated by the evanescent radiation produced by the body and confined near its surface. The absence of an evanescent contribution associated to the surrounding walls results in an atomic state which, even for large atom-body distances, does not tend to a thermal state at temperature $T_\text{W}$. Far from the surface, the atom sees only propagative contributions to the radiation (in other words, the atom is always far from the surrounding walls), and its state remains dependent on the temperature and properties of the body.

\subsection{Three-level atom}\label{par:NAqutrit}

We now turn our attention to the case of a three-level system. As we will see, in this configuration the presence of two different effective temperatures (each one associated to one of the two allowed atomic transitions) allows to produce a wider variety of physical situations, and in particular the population inversion of the two lowest-energy states.

\begin{figure}[h]
\includegraphics[width=0.48\textwidth]{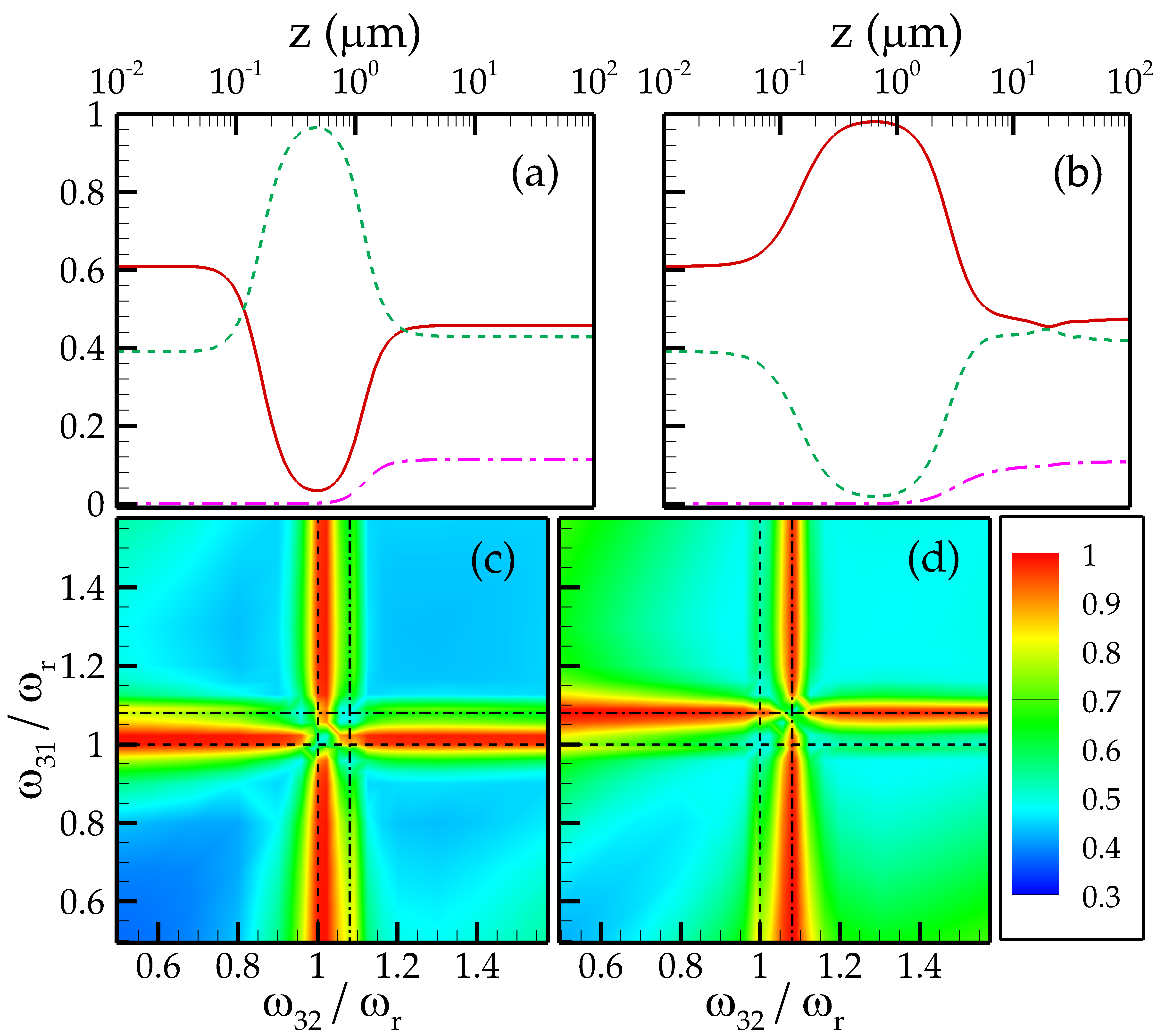}
\caption{(color online). Steady populations (red solid line for $\rho_{11}(\infty)$, green dashed line for $\rho_{22}(\infty)$, purple dot-dashed line for $\rho_{33}(\infty)$) as a function of $z$ for $\delta=0.01\,\mu$m (a) and $\delta=2\,\mu$m (b), being $(T_\text{W},T_\text{M})=(300,50)\,$K, $\omega_{32}=1.02\,\omega_r$ and $\omega_{31}=\omega_p$. Density plot of purity as a function of the two frequencies $\omega_{32}$ and $\omega_{31}$ for $z=0.47\,\mu$m, $\delta=0.01\,\mu$m (c) and $\delta=2\,\mu$m (d). The black dotted lines correspond to $\omega_r$ and $\omega_p$ and highlight the zones where high values of purity are obtainable.}
\label{Fig6}\end{figure}

In Fig. \ref{Fig6}(a)-(b) we plot the steady populations as a function of $z$ for two different values of $\delta$, choosing $(T_\text{W},T_\text{M})=(300,50)\,$K, $\omega_{32}=1.02\,\omega_r$ and $\omega_{31}=\omega_p$. The plots show the occurrence of population inversion between the states $\ket{1}$ and $\ket{2}$. It is also shown that by varying only the slab thickness, for values of $z$ around $0.5\,\mu$m one can pass from a steady state with almost all the population in $\ket{2}$ to another one with almost all the population in $\ket{1}$. This kind of behavior corresponds to high values of purity which quantifies how pure the atomic state is \cite{BookBreuer}. This is defined for an arbitrary three-level system by
\begin{equation}\Pi (\rho)=\mathrm{Tr}[\rho^2]=\sum_i\rho_{ii}^2+2\bigl(|\rho_{12}|^2+|\rho_{13}|^2+|\rho_{23}|^2\bigr).\end{equation}
In the steady configuration the state is diagonal and high values of purity are reached when $\rho_{11}(\infty)$ or $\rho_{22}(\infty)$ tends to 1 (assuming that $\rho_{33}(\infty)$ remains much smaller than both $\rho_{11}(\infty)$ and $\rho_{22}(\infty)$). In Fig. \ref{Fig6}(c)-(d) the density plot of purity is represented by varying the two frequencies $\omega_{32}$ and $\omega_{31}$ for $z=0.47\,\mu$m and $\delta=0.01\,\mu$m in (c) and $\delta=2\,\mu$m in (d). The black dotted lines correspond to $\omega_r$ and $\omega_p$, pointing out that high values of purity can be obtained only when at least one frequency is close to slab resonances.

\begin{figure}[h]
\includegraphics[width=0.49\textwidth]{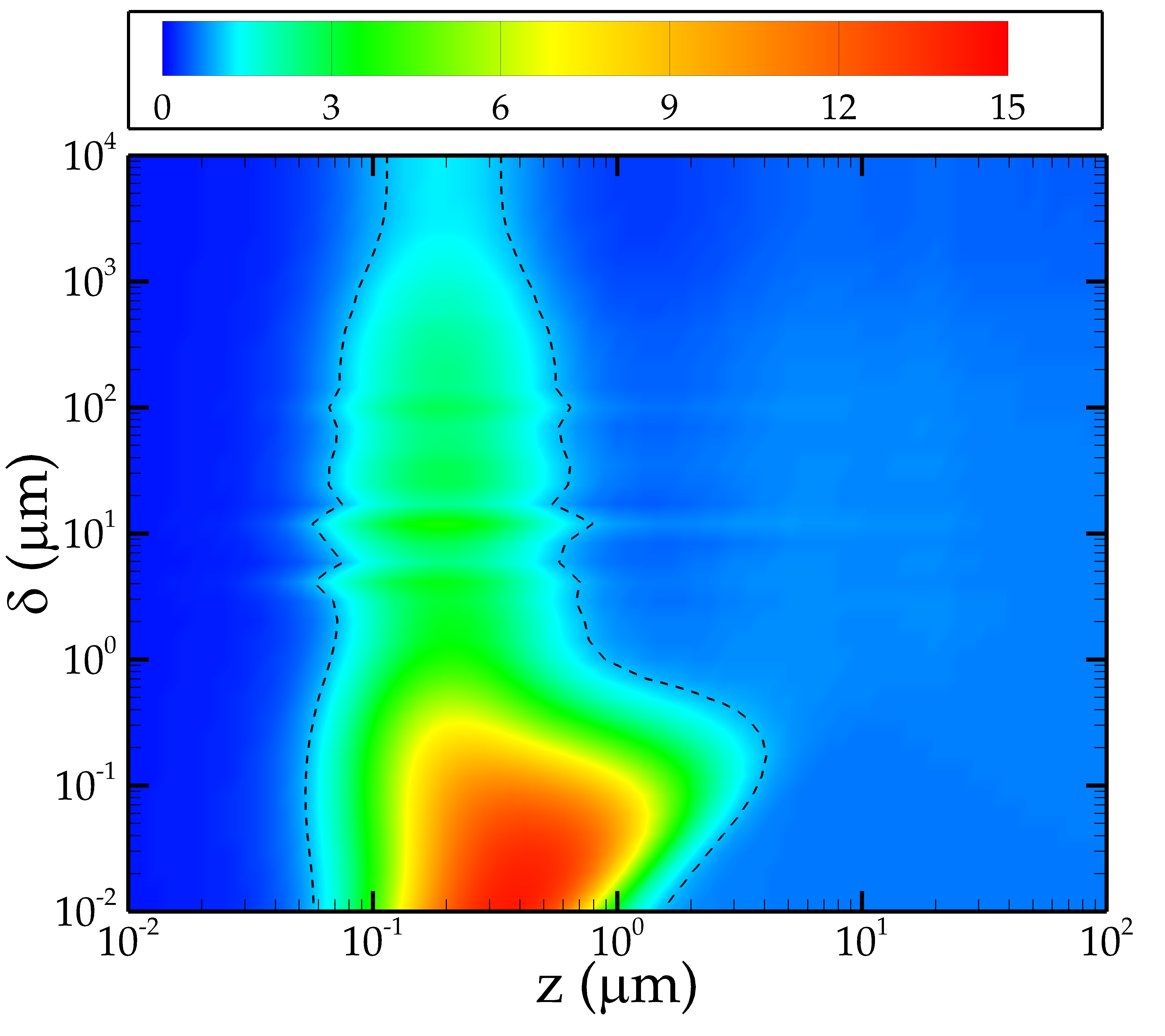}
\caption{(color online). Density plot of $\rho_{22}(\infty)/\rho_{11}(\infty)$ as a function $z$ and $\delta$ for $(T_\text{W},T_\text{M})=(500,100)\,$K, $\omega_{32}=1.02\,\omega_r$ and $\omega_{31}=1.57\,\omega_r$. The dotted line, corresponding to $\rho_{22}(\infty)/\rho_{11}(\infty)=1$, highlights a large regions of occurrence of  inversion of population ordering as well as a strong dependence on $z$ and $\delta$.}
\label{Fig7}\end{figure}

In order to stress the occurrence of inversion of population ordering and its dependence on the atom-slab distance and the slab thickness, in Fig. \ref{Fig7} we present a density plot of the ratio $\rho_{22}(\infty)/\rho_{11}(\infty)$ as a function of $z$ and $\delta$ for $(T_\text{W},T_\text{M})=(500, 100)\,$K, $\omega_{32}=1.02\,\omega_r$ and $\omega_{31}=1.57\,\omega_r$. The dotted line corresponds to a ratio equal to 1, that is when the two populations coincide. This plot points out a strong dependence of the ratio on the two involved parameters, with regions where the ratio is larger than one characterized by inversion of population. This indeed occurs even if one of the two frequencies is far from $\omega_r$.

\begin{figure}[h]
\includegraphics[width=0.49\textwidth]{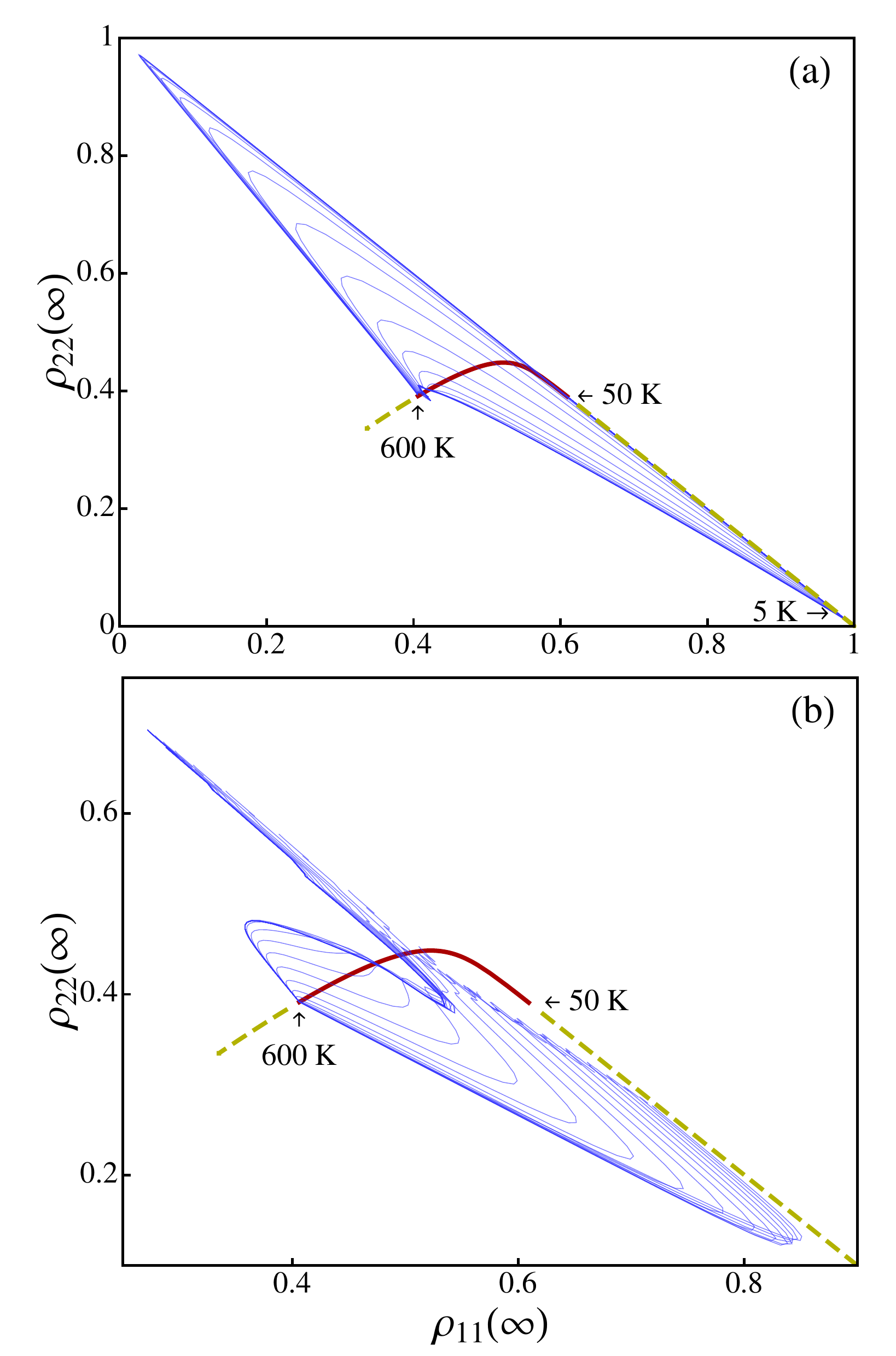}
\caption{(color online). Region occupied by the couples $(\rho_{11}(\infty),\rho_{22}(\infty))$, for $z$ varying in $[0.01,100]\,\mu$m and $\delta$ in $[0.01\,\mu$m,$1\,$cm], for $\omega_{32}=1.02\,\omega_r$ and $\omega_{31}=\omega_p$. The red curve represents the thermal states between $T_\text{W}$ and $T_\text{M}$ while the yellow dashed curve corresponds to temperatures exceeding the ones involved in the system. For each value of $\delta$ we represent a continuous curve by varying $z$. In (a) the chosen temperatures are $(T_\text{W},T_\text{M})=(600,50)\,$K while in (b) we have $(T_\text{W},T_\text{M})=(50,600)\,$K.}
\label{Fig8}\end{figure}

In Fig. \ref{Fig8}(a) we trace the surface of points $(\rho_{11}(\infty),\rho_{22}(\infty))$ by varying $z$ in $[0.01,100]\,\mu$m and $\delta$ in [$0.01\,\mu$m,$1\,$cm] for $(T_\text{W},T_\text{M})=(600,50)\,$K, $\omega_{32}=1.02\,\omega_r$ and $\omega_{31}=\omega_p$. The region covered by the steady states is compared with the curve of thermal states between $T_\text{W}$ and $T_\text{M}$ (red continuous line) and outside them (yellow dashed curve). The plot proves that a large variety of steady states can be obtained. In particular, for $\delta\gtrsim3\,\mu$m (semi-infinite slab) and $z=0.54\,\mu$m, the steady state is almost a thermal state at $T\simeq5\,$K (this value is obtained by looking for the closest thermal state by using the trace-norm distance). This cooling mechanism is connected to the behavior of the effective temperatures, which for these values of $z$ and $\delta$ are $T_{\mathrm{eff}}^{(32)}\simeq145\,$K and $T_{\mathrm{eff}}^{(31)}\simeq59\,$K. We remark here that the cooling concerns only
the internal atomic degrees of freedom and not the external motion. In Fig. \ref{Fig8}(b) the same plot is presented by inverting the two temperatures $(T_\text{W},T_\text{M})=(50,600)\,$K. The resulting shape of the surface is completely different, showing that the behavior of the steady state is not symmetric for exchange of the two temperatures. In particular, for large values of $z$ the steady state remains quite far from the thermal state at $T_\text{W}$.

\begin{figure}[h]
\includegraphics[width=0.47\textwidth]{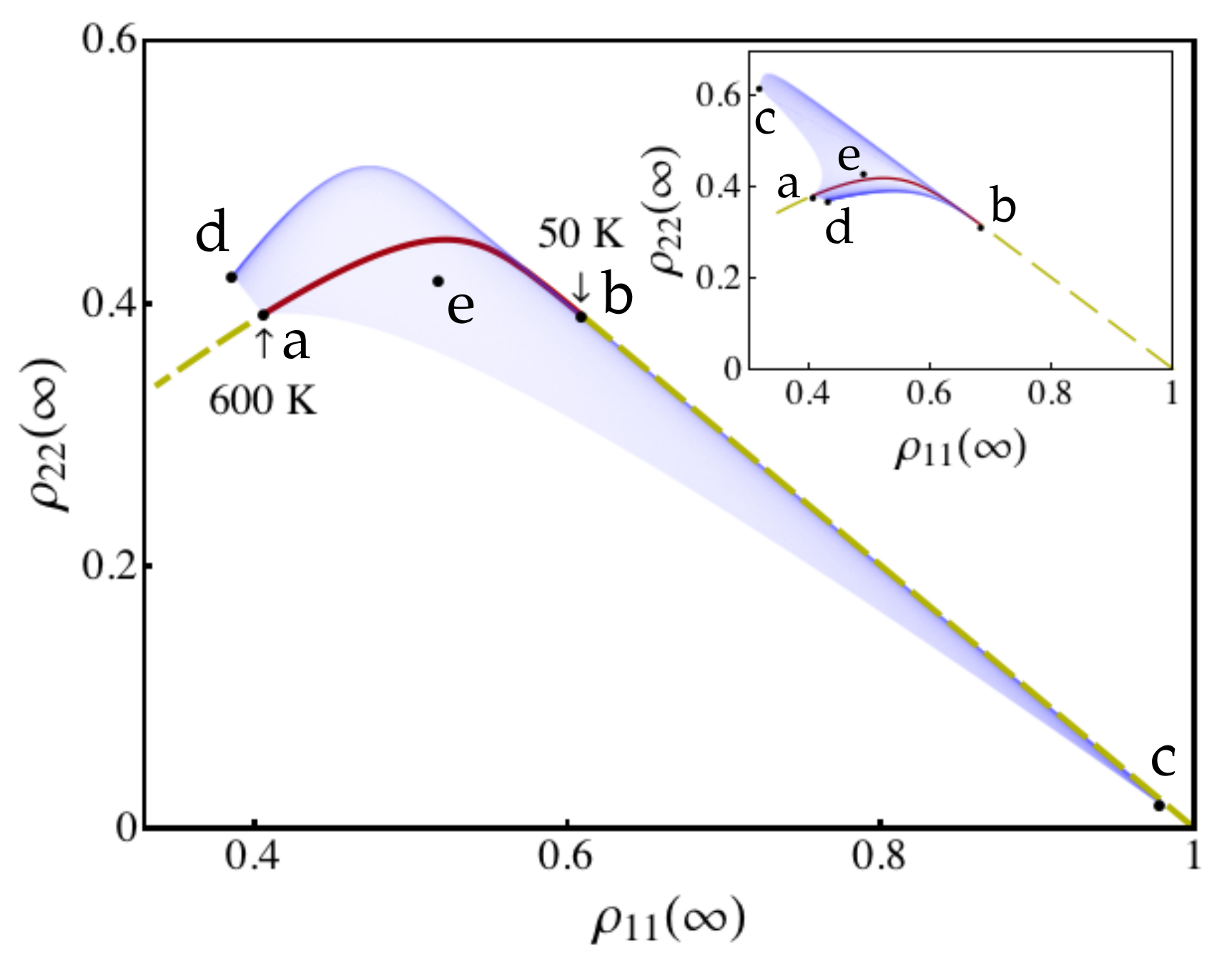}
\caption{\label{Fig11}\footnotesize Points ($\rho_{11}(\infty)$,$\rho_{22}(\infty)$) for all possible couples of temperatures ($T_\mathrm{W}$,$T_\mathrm{M}$) in the interval $[50\,$K,$600\,$K$]$ for $\delta=300\,\mu$m (semi-inifinite slab) and $z=1\,\mu$m. Thermal states are represented by red solid line ($50\,\mathrm{K}<T<600\,\mathrm{K}$) and yellow dashed line ($T<50\,$K and $T>600\,$K). In the main plot, $\omega_{32}=1.02\,\omega_r$ and $\omega_{31}=\omega_p$. In the inset, $\omega_{32}=0.9\,\omega_r$ and $\omega_{31}=\omega_r$. Darker zones correspond to a higher density of states. By moving along the border of the surface we have the following variations of temperature. From the comparison with the inset one sees that in correspondence with the same change of temperatures, different trajectories are obtained. From `a' ($(T_\text{W},T_\text{M})=(600,600)\,$K) to `c' ($(T_\text{W},T_\text{M})=(600,50)\,$K), $T_\text{W}$ is fixed while $T_\text{M}$ goes from 600\,K to 50\,K. From `c' to `b' ($(T_\text{W},
T_\text{M})=(50,50)\,$K),
 $T_\text{M}$ is fixed while $T_\text{W}$ goes from 600\,K to 50\,K.  From `b' to `d' ($(T_\text{W},T_\text{M})=(50,600)\,$K), $T_\text{W}$ is fixed while $T_\text{M}$ goes from 50\,K to 600\,K. From `d' to `a', $T_\text{M}$ is fixed while $T_\text{W}$ goes from 50\,K to 600\,K. `e' ($(T_\text{W},T_\text{M})=(250,200)\,$K) represents a point internal to the surface.}
\end{figure}

In Fig. \ref{Fig11} we show that, by varying only $T_\mathrm{W}$ and $T_\mathrm{M}$ in the interval $[50,600]\,$K while keeping all the other parameters fixed, large regions in the space of states can be explored. Different shapes are realized when one varies for example $\omega_{31}$ and $\omega_{32}$, as shown in the inset. Figure \ref{Fig11} confirms that non-equilibrium configurations provide new tools to realize a large variety of non-trivial steady states.

\subsection{Gold slab}\label{par:metal}

In this subsection we briefly discuss the qualitative modifications to the atomic dynamics induced by the replacement of GaAs with a metallic surface. To this aim we consider a slab made of gold, whose dielectric permittivity $\epsilon (\omega)$ is described using a Drude model
\begin{equation}\epsilon(\omega)=1-\frac{\omega_{\textrm{pl}}^2}{\omega^2+i\omega\gamma},\end{equation}
characterized by a plasma frequency $\omega_{\textrm{pl}}=137.2\times10^{14}\,\mathrm{rad}\,\text{s}^{-1}$ and where $\Gamma=0.4059\times10^{14}\,\mathrm{rad}\,\text{s}^{-1}$. This model implies a surface plasmon resonance at $\omega_p=96.987\times10^{14}\,\mathrm{rad}\,\text{s}^{-1}$. Both $\omega_\text{pl}$ and $\omega_p$ correspond to temperatures much higher than a typical room temperature of 300\,K. However, in order to make effects out of thermal equilibrium emerge, we prefer to limit our analysis at frequencies around room temperature, that is around $\omega_\text{R}=300k_B/\hbar\simeq0.392\times10^{14}\,\mathrm{rad}\,\text{s}^{-1}$. In this range of frequencies we observe much less richness in the behavior of the quantities of interest as a function of the various parameters.

\begin{figure}[h]
\includegraphics[width=0.50\textwidth]{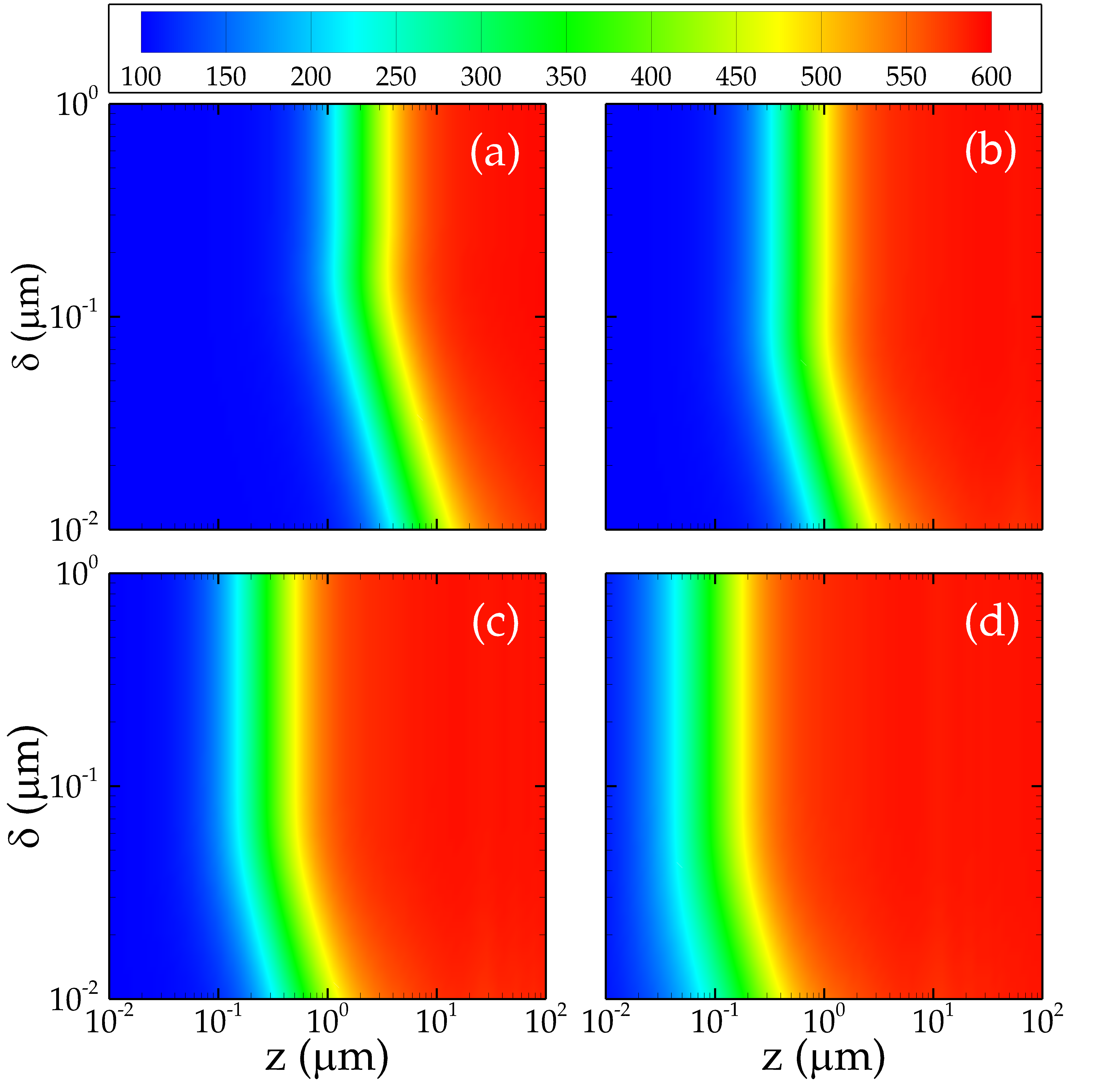}
\caption{(color online). Atom in front of a gold slab. Density plot of $T_{\mathrm{eff}}^{(nm)}$ as a function of $z$ and $\delta$ for four different frequencies for $(T_\text{W},T_\text{M})=(600,100)\,$K. The chosen frequencies are: $0.1\,\omega_\text{R}$ (a), $0.5\,\omega_\text{R}$ (b), $\omega_\text{R}$ (c), and $2.5\,\omega_\text{R}$ (d).}
\label{Fig13}\end{figure}

Analogously with what presented in Fig. \ref{Fig4}, we plot in Fig. \ref{Fig13} the effective temperature $T_{\mathrm{eff}}^{(nm)}$ as a function of $z$ and $\delta$ for four different frequencies and $(T_\text{W},T_\text{M})=(600,100)\,$K. By varying $\omega_{nm}$ around $\omega_r$ the only visible effect is a global horizontal shift of the plot, corresponding to the fact that by increasing the frequency the region where the effect of the slab dominates decreases. This effect is connected to the rapid decay with $\omega$ of the imaginary part of the dielectric permittivity, which in turn reduces the contribution of the term $\mathbf{D}(\omega_{nm})$ (the only one diverging at smaller distances) to $\alpha_\mathrm{M}(\omega_{nm})$ (see Eq. \eqref{alphaE and alphaM} and Appendix \ref{par:integral}).

\begin{figure}[h]
\includegraphics[width=0.50\textwidth]{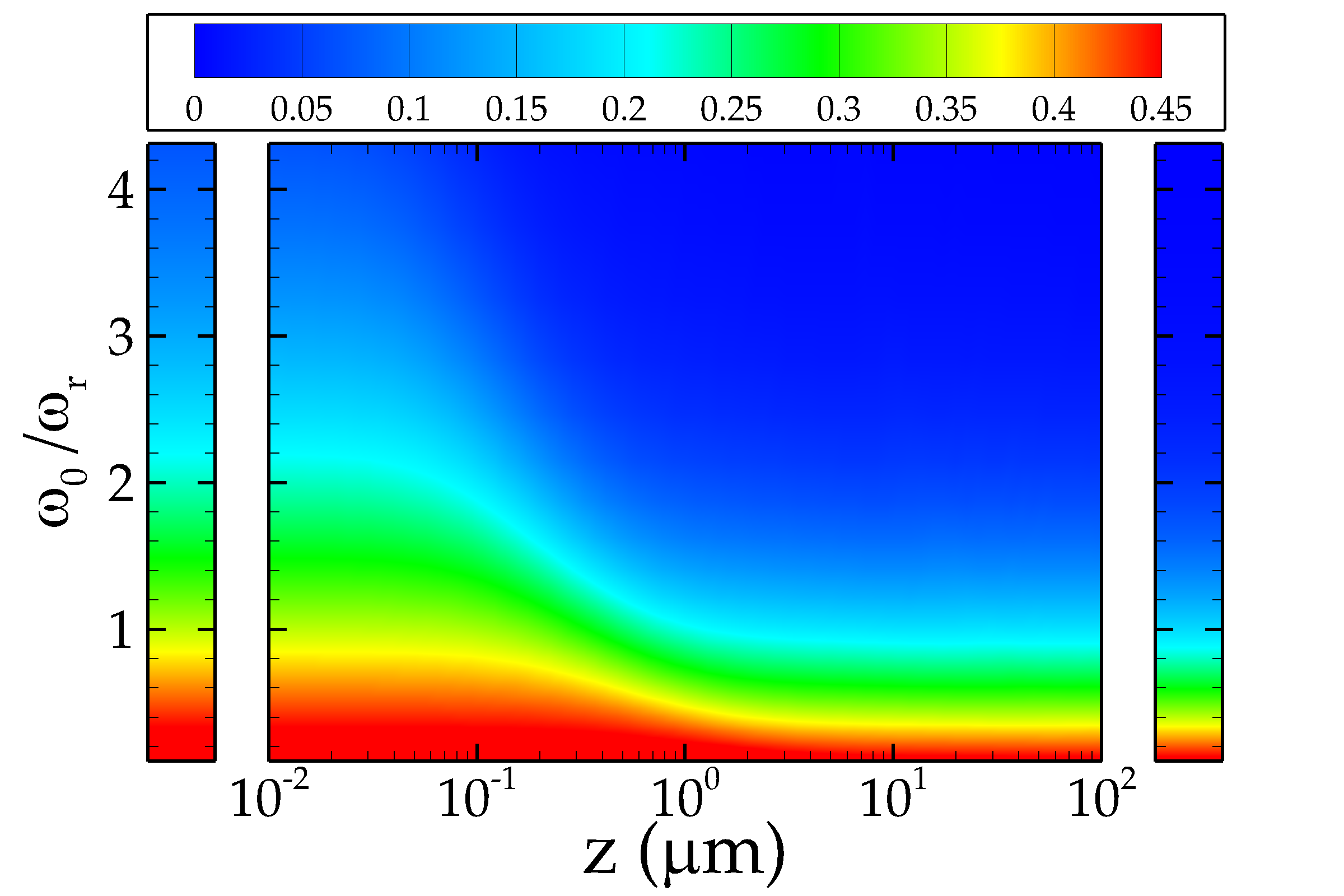}
\caption{(color online). Atom in front of a gold slab. Density plot of the excited state population $\rho_{22}(\infty)$ as a function of $z$ and $\omega_0$ for $\delta=1\,$cm (semi-infinite slab). The temperatures are $(T_\text{W},T_\text{M})=(200,500)\,$K. On the left (right) of the main part, the density plot at thermal equilibrium at 500\,K (200\,K) is given.}
\label{Fig14}\end{figure}

In the case of a two-level system, in analogy with Fig. \ref{Fig5}, we plot in Fig. \ref{Fig14} the excited-state population $\rho_{22}(\infty)$ as a function of $z$ and $\omega_0$ for $\delta=1\,$cm (semi-infinite slab). The plot highlights a smoother dependence of $\rho_{22}(\infty)$ on $z$. Moreover, we see that in the case of a metal thermal-equilibrium values at $T_\text{W}$ are always recovered for large $z$. This is technically due to the fact that in Eq. \eqref{alphaE and alphaM} the contribution $\mathbf{B}(\omega_{nm})$ has always values close to 1 so that in the zone where it dominates (that is asymptotically in $z$), it makes only $\alpha_\mathrm{W}(\omega_{nm})$ contribute and thus the slab temperature irrelevant.

\begin{figure}[h]
\includegraphics[width=0.49\textwidth]{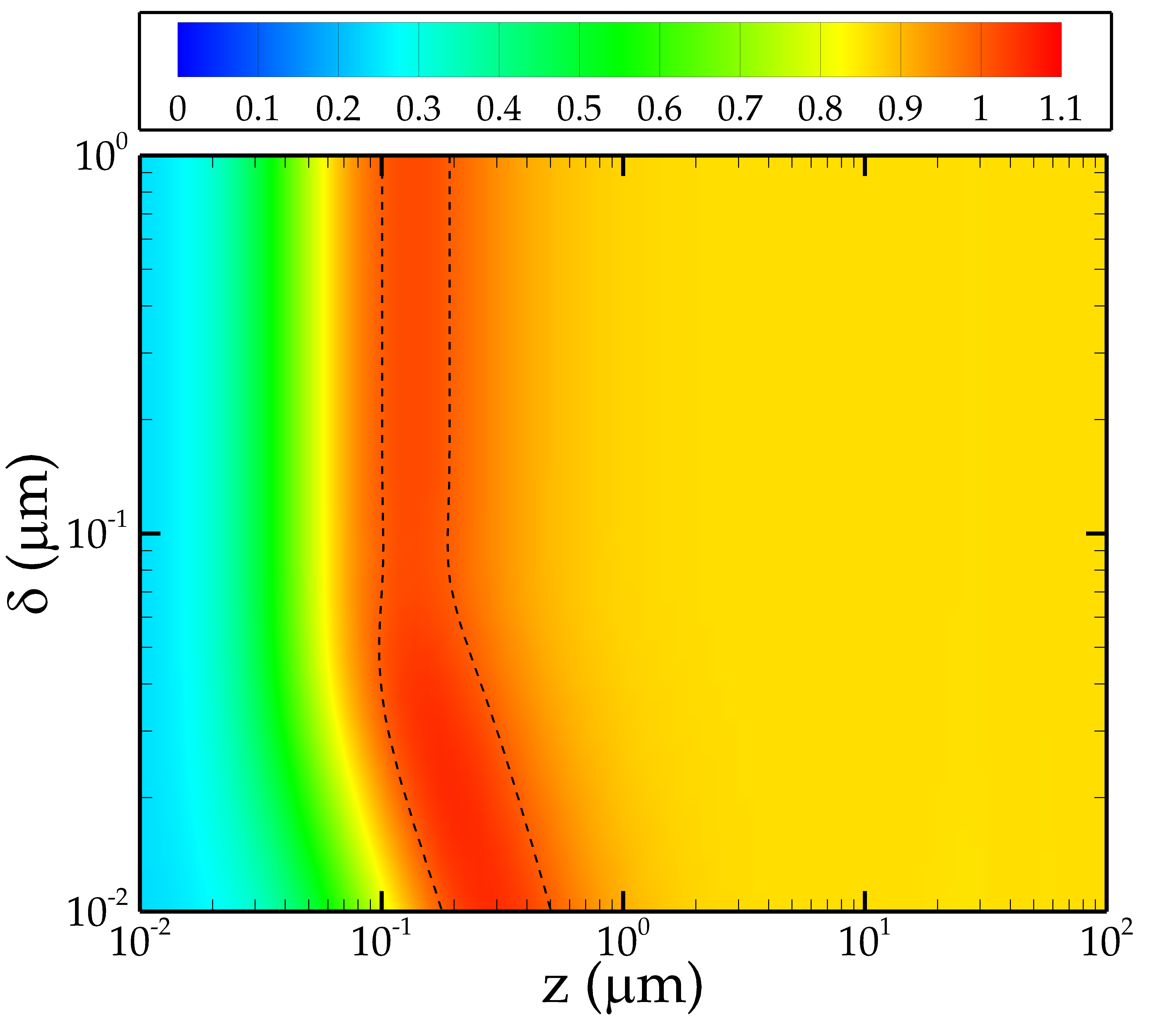}
\caption{(color online). Atom in front of a gold slab. Density plot of $\rho_{22}(\infty)/\rho_{11}(\infty)$ as a function $z$ and $\delta$ for$(T_\text{W},T_\text{M})=(600,60)\,$K, $\omega_{32}=0.704\,\omega_\text{R}$ and $\omega_{31}=\omega_\text{R}$. We observe a weak dependence on $z$ and $\delta$, with small regions of occurrence of inversion of population ordering.}
\label{Fig15}\end{figure}

In the case of a three-level system, in analogy with Fig. \ref{Fig7}, we plot in Fig. \ref{Fig15} the ratio $\rho_{22}(\infty)/\rho_{11}(\infty)$ as a function $z$ and $\delta$ for $(T_\text{W},T_\text{M})=(600,60)\,$K, $\omega_{32}=0.704\,\omega_\text{R}$ and $\omega_{31}=\omega_\text{R}$. Black dotted lines indicate where the ratio is equal to 1, that is when the two populations coincide. The comparison between this figure and Fig. \ref{Fig7} shows that in this case the dependence of the ratio on the two involved parameters is less rich and inversion of population is reached in a small region of the $(z,\delta)$ plane, with $\rho_{22}(\infty)$ only slightly larger than $\rho_{11}(\infty)$.

\section{Conclusions}\label{par:Conclusions}

In this paper we have reported a detailed investigation of the dynamics of an elementary quantum system placed close to an arbitrary body whose temperature $T_\mathrm{M}$ is kept fixed and different from that of the surrounding walls $T_\mathrm{W}$. We have derived a suitable master equation for the atomic dynamics in such stationary environments out of thermal equilibrium, providing closed-form analytic expressions for the transition rates governing the dynamics in terms of scattering matrices of the body. We have pointed out relevant differences with respect to the case of thermal equilibrium, with the steady state now dependent on a complex interplay between the atom-body distance and the geometrical and material properties of the body. The overall dynamics can be readily interpreted in terms of effective temperatures associable to each allowed transition. After treating the general case of a $N$-levels atom, the case of two- and there-level
atoms have been discussed in detail. In the first case, the atom thermalizes to a thermal state at a temperature comprised between $T_\mathrm{M}$ and $T_\mathrm{W}$, while in the second case the steady state is not thermal in general and the steady population may exceed the equilibrium values at $T_\mathrm{M}$ and $T_\mathrm{W}$. This effect, emerging out of equilibrium, allows one to strongly manipulate the steady state, realizing inversion of population ordering and cooling of internal atomic temperature at values external to the interval between the involved temperatures. We have then specialized our analysis to the case in which the body is a slab of finite thickness, deriving explicit expressions for the transition rates. We have provided a detailed numerical investigation for both the cases when the slab is made of a dielectric and a metal, pointing out that dielectric configurations present a richer variety in controlling the atomic dynamics. The
slab thickness regulates the distances from the slab at which the evanescent contribution of the field emitted by the slab dominates the atomic dynamics which, in this limit, becomes similar to an equilibrium dynamics at $T_\mathrm{M}$. Moving the atom far enough from the slab, the effect of $T_\mathrm{W}$ becomes relevant. This effect depends on the relative value of the frequency of the transition with respect to the slab resonances. As a consequence, for each transition the influence of the two temperatures can be also quite different, allowing a strong manipulation of the atomic dynamics by varying $T_\mathrm{M}$ and $T_\mathrm{W}$.

The results reported in this paper could be of interest for experimental investigations in the absence of thermal equilibrium involving real or artificial atoms (such as quantum dots). Our work suggests that similar effects will be present also in the case of more atoms trapped close to a substrate \cite{ObrechtPRL07} or traveling in an atomic beam \cite{HindsPRL93} in proximity of a surface. For quantum dots one can also consider mechanical manipulation of the dot \cite{Saidi09}.

\begin{acknowledgments}
The authors acknowledge G. Cassabois and  E. Rousseau for useful discussions. BB and MA acknowledge financial support from the Julian Schwinger Foundation. MA acknowledges the Labex NUMEV.  \end{acknowledgments}

\appendix

\section{Approximations}\label{par:discussions on approximations}

In this appendix we briefly discuss the approximations used in the derivation of Sec. \ref{par:derivation} \cite{BookBreuer}.

The first approximation consists of a weak-coupling assumption which allows to expand the exact equation of motion for the density matrix up to second order in the coupling constant. Together with the condition $\rho_{\text{tot}}(t)\simeq\rho(t)\rho_B$ this leads to the Born approximation of the master equation. The second approximation is the Markov approximation in which the quantum master equation is made local in time: the density matrix $\rho(t)$, in principle depending on $\rho(s)$ for any $s\leq t$, is assumed to depend only on the density matrix at the same time $t$. Furthermore, the master equation is made independent of the initial state of the system. This Born-Markov limit is valid if the bath correlation time $\tau_B$ is small compared to the relaxation time of the system $\tau_R$, that is $\tau_B\ll\tau_R$. Finally, in the rotating wave approximation, rapidly oscillating terms proportional to $\exp[i(\omega'-\omega)t]$ for $\omega'\neq\omega$ are neglected, ensuring that the quantum
master equation is in Lindblad form. The corresponding condition is that the inverse frequency differences involved in the problem are small compared to the relaxation time of the system, that is $\tau_A\simeq|\omega'-\omega|^{-1}\ll\tau_R$. The master equation given in Eq. \eqref{master equation} can be used only for values of the various parameters (coupling constants, temperatures, atom frequencies, etc.) such that the conditions stated above are satisfied.

We first estimate $\tau_R$ for a  given transition of frequency $\omega_{nm}$. For a two-level atom, we argue from Eq. \eqref{rhoevolution} that it is given by
\begin{equation}\tau_R\simeq\bigl[\gamma(\omega_{nm})\bigr]^{-1}=\bigl[\Gamma(-\omega_{nm})+\Gamma(\omega_{nm})\bigr]^{-1}.\end{equation}
For an atom with more than two levels, $\tau_R$ will be a complex function of all the various transition rates. For our numerical estimation, we consider the typical case of dipole momentum of the order of $10^{-29}$\,C\,m. In order to test the validity of  the Born-Markov approximation, we focus on the worst case. The smallest values of $\tau_R$ are obtained for $\omega_{nm}=\omega_p$, for small values of $z$ (which in our numerical analysis start from 0.01\,$\mu$m) and for large values of $\delta$. Let us fix the temperatures as $T_\text{max}=600$\,K, $T_{\text{min}}=100$\,K and $\delta=$1\,cm. The smallest value of $\tau_R$, obtained for $z=0.01\,\mu$m and $T_{\text{M}}=T_{\text{max}}$, is $\tau_R\simeq4\times10^{-11}$\,s. In this limit, the results at equilibrium ($T_{\text{W}}=T_{\text{max}}$) and out of equilibrium ($T_{\text{W}}=T_{\text{min}}$) are equal (only the slab temperature is relevant). By increasing $z$, $\tau_R$ rapidly increases. At $z\simeq0.04\,\mu$m we have $\tau_R\simeq2\
times10^{-9}
$\,s. Also by slightly moving $\omega_{nm}$ around $\omega_p$ we observe a rapid increase of $\tau_R$. For example at $\omega_{nm}\simeq1.04\,\omega_p$ and $z\simeq0.01\,\mu$m, $\tau_R\simeq4\times10^{-9}$\,s. At $\omega_{nm}=\omega_p$ and large values of $z$, we find very large times being $\tau_R\simeq0.005$\,s for $T_{\text{W}}=T_{\text{max}}$. Similar considerations and numerical values are found for the three-level case in a $\Lambda$ configuration analyzed along the paper.

We now estimate $\tau_B$ by looking at the decay time of $\langle E_i(\mathbf{R},s)E_j(\mathbf{R},0)\rangle$. This means computing (compare with Eq. \eqref{gammaijdef})
\begin{equation}\label{correlation function}\langle E_i(\mathbf{R},s)E_j(\mathbf{R},0)\rangle
=\frac{\hbar^2}{2 \pi} \int_{-\infty}^{+\infty} d\omega \: e^{-i\omega s} \gamma_{ij}(\omega).\end{equation}
We can thus numerically estimate $\tau_B$ and compare it with $\tau_R$. In absence of the body the  decay time $\tau_B$ is fixed by the thermal fluctuations giving times for  $\tau_B$ of the order of $10^{-13}$\,s. In presence of the body,
$\tau_B$ has a weak dependence on $z$ and the two temperatures, and assumes values of the order of $(1-10)\times10^{-12}$\,s. These values are almost always much smaller than $\tau_R$, becoming close only for $z\simeq0.01\,\mu$m. Once again, close to the slab only the slab temperature is relevant, so that the analyses at equilibrium and out of equilibrium give analogous results. Being the condition $\tau_B\ll\tau_R$ always satisfied for values of the dipole momentum small enough, then the Born-Markov approximation is well satisfied in our system.

With regards to the rotating wave approximation, in the case of a two-level system $\tau_A\sim(2\omega_{nm})^{-1}$, so that one has to check if $(2\omega_{nm})^{-1}\ll\bigl[\Gamma(-\omega_{nm})+\Gamma(\omega_{nm})\bigr]^{-1}$. In our system, the frequencies investigated are of the order of $(0.05 -1.5)\times10^{14}$\,rad\,s$^{-1}$, so that $ \tau_A$ is of the order of $(0.3-10)\times10^{-14}$\,s. These values are typically much smaller than $\tau_R$, which, as discussed before, in the worst case is of the order of $4\times10^{-11}$\,s. Similar considerations hold for the three-level atom considered in the paper.

\section{Integrals $\mathbf{B}(\omega)$, $\mathbf{C}(\omega)$ and $\mathbf{D}(\omega)$}\label{par:integral}

This section is devoted to a brief discussion of the dependence of $\mathbf{B}(\omega)$, $\mathbf{C}(\omega)$ and $\mathbf{D}(\omega)$ (defined in Eq. \eqref{integrals}) on the atom-slab distance $z$ and the slab thickness $\delta$.

Let us start with the contribution $\mathbf{B}(\omega)$, the only one appearing both in $\alpha_\text{W}$ and $\alpha_\text{M}$. As remarked before, it does not depend on $z$ but only on the thickness $\delta$. Using Eq. \eqref{integral A} and noting that the sum $|\rho_{p}(k,\omega)|^2+|\tau_{p}(k,\omega)|^2$ belongs to $[0,1]$, we have as an immediate result that $0<[\mathbf{B}(\omega)]_i<1$ for $i=x,y,z$. Moreover, for $\delta\to+\infty$ we have $\tau_p(k,\omega)\to0$ and we can replace $\rho_p(k,\omega)$ with the bulk Fresnel coefficient $r_p(k,\omega)$, obtaining
\begin{equation}\mathbf{B}(\omega)=\frac{3c}{4\omega}\sum_p\int_0^{\frac{\omega}{c}}\frac{k\,dk}{k_z}\mathbf{M}_p^+(k,\omega)|r_p(k,\omega)|^2.\end{equation}
Finally, for $\delta=0$ (namely in the absence of the slab) we have $\rho_p(k,\omega)=0$ and $\tau_p(k,\omega)=1$ so that $[\mathbf{B}(\omega)]_i=1$. We observe that in the same limit we have $\mathbf{C}(\omega)=0$ and $\mathbf{D}(\omega)=0$, so that $\alpha_\text{M}(\omega)=0$, i.e. the dependence on the properties of the slab naturally disappears.

Let us now discuss the properties of $\mathbf{C}(\omega)$. We first observe that $\mathbf{C}(\omega)$ is finite in the limit $z\to0$, as a consequence of Lebesgue dominated convergence theorem. As for the limit $z\to+\infty$, we set $\tilde{k}=\frac{ck}{\omega}$, $\tilde{z}=\frac{\omega z}{c}$ and we make the change of variable $\sqrt{1-\tilde{k}^2} \to s$. The integral $\mathbf{C}(\omega)$ is reduced to the form 
\begin{equation}\int_0^1g(s)e^{2is\tilde{z}}ds,\end{equation}
for some smooth function $g(s)$. An integration by parts then shows that $\mathbf{C}(\omega) \to 0$ as $\tilde{z} \to +\infty$.

The asymptotic behavior of  $\mathbf{D}(\omega)$ for $z\to +\infty$  is obtained by noting that the integrals involved are, after the change of variable $\sqrt{\tilde{k}^2-1} \to s$, of the form
\begin{equation}\int_0^{+\infty} g(s) e^{-2s\tilde{z}}\,ds,\end{equation}
where the function $g(s)$ has a polynomial behavior near $s=0$, i.e. $g(s)\sim{\cal{G}}\,s^n$ for some non negative integer $n$. We can then use the Lebesgue dominated convergence theorem to show that near $\tilde{z}\to+\infty$ one has
\begin{equation}\int_0^{+\infty} g(s) e^{-2s\tilde{z}}\,ds \sim {\cal{G}}\frac{n!}{2^{n+1}}\frac{1}{\tilde{z}^{n+1}}.\end{equation}
It now remains to specify the value of the constant ${\cal{G}}$  for each component of $\mathbf{D}$. The components $x$ and $y$ of $\mathbf{D}(\omega)$ are equal and are denoted $D_{xy}$. In particular, the terms connected to the two polarizations $\text{TE}$ and $\text{TM}$ are denoted $D_{xy}^{(\text{TE})}$ and $D_{xy}^{(\text{TM})}$ respectively. They are given by
\begin{equation}\label{integralnv}
\begin{split}D_{xy}^{(\text{TE})}&=\frac{3}{4}\int_0^{\infty }e^{-2 s \tilde{z}}\Ima\bigl(\rho_1 (s,\delta,\epsilon)\bigr)\,ds,\\
D_{xy}^{(\text{TM})}&=\frac{3}{4}\int_0^{\infty }e^{-2 s \tilde{z}}s^2 \Ima\bigl(\rho_2(s,\delta,\epsilon)\bigr)\,ds.\end{split}\end{equation}

For the $z$ component $D_{z}$ of $\mathbf{D}(\omega)$, we have
\begin{equation}\label{integralnvz}\begin{split}D_{z}^{(\text{TE})}&=0,\\
D_{z}^{(\text{TM})}&=\frac{3}{2}\int_0^{\infty }e^{-2 s \tilde{z}}\left(1+s^2\right)\Ima\bigl(\rho_2(s,\delta,\epsilon)\bigr)ds.\end{split}\end{equation}

The following asymptotic expansions near $s=0$ hold
\begin{equation}\begin{split}\Ima\bigl(\rho_1(s,\delta,\epsilon)\bigr)&\sim-2s\Ima\bigl(f(\delta,\epsilon)\bigr),\\
\Ima\bigl(\rho_2(s,\delta,\epsilon)\bigr)&\sim-2s\Ima\bigl(\epsilon\,f(\delta,\epsilon)\bigr),\end{split}\end{equation}
where $f(\epsilon,\delta)=\frac{\cot\left(\delta\sqrt{\epsilon-1}\right)}{\sqrt{\epsilon-1}}$. We then obtain for $\tilde{z}\to \infty$
\begin{center}
\begin{tabular}{|c|c|c|c|c|}
\hline
 &$D_{xy}^{(\text{TE})}$&$D_{xy}^{(\text{TM})}$&$D_{z}^{(\text{TM})}$\\
\hline
\rule[-4mm]{0mm}{1cm}
$\tilde{z}=\frac{\omega z}{c}\to \infty$ &$\frac{-3\Ima\left(f(\epsilon,\delta)\right)}{8 \tilde{z}^2}$&$\frac{-9\Ima\left(\epsilon f(\epsilon,\delta)\right)}{16 \tilde{z}^4}$&$\frac{-3 \Ima\left(\epsilon f(\epsilon,\delta)\right)}{4 \tilde{z}^2}$ \\
\hline
\end{tabular}
\end{center}

Finally, we evaluate the behavior of the integrals with respect to $\delta$ near $0$ and $+\infty$. Denoting $\tilde{\delta}=\frac{\omega\delta}{c}$, we have $\cot\left(\delta\sqrt{\epsilon-1}\right)\to\frac{1}{\delta \sqrt{\epsilon-1}}$ for $\delta \to 0$ and $\cot\left(\delta\sqrt{\epsilon-1}\right)\to-i$ for $\delta\to\infty$, so that
\begin{center}
\begin{tabular}{|c|c|c|c|c|}
\hline
$\tilde{z}\to \infty$&$D_{xy}^{(\text{TE})}$&$D_{xy}^{(\text{TM})}$&$D_{z}^{(\text{TM})}$\\
\hline
\rule[-4mm]{0mm}{1.2cm}
$\frac{\omega\delta}{c}\!\to\! \infty $ &$\frac{3 }{8 \tilde{z}^2}\!\Rea\!\left(\!\frac{1}{\sqrt{\epsilon-1}}\!\right)$&$\frac{9}{16 \tilde{z}^4}\!\Rea\! \left(\!\frac{\epsilon}{\sqrt{\epsilon-1}}\!\right)$&$\frac{3 }{4 \tilde{z}^2}\!\Rea\!\left(\!\frac{\epsilon}{\sqrt{\epsilon-1}}\!\right)$\\
\rule[-4mm]{0mm}{1.2cm}
$\frac{\omega\delta}{c}\!\to\! 0$&$\frac{-3 }{8 \tilde{z}^2 \tilde{\delta}} \!\Ima\! \left(\!\frac{1}{\sqrt{\epsilon-1}}\!\right)$&$\frac{-9}{16 \tilde{z}^4 \tilde{\delta}} \!\Ima\! \left(\!\frac{\epsilon}{\sqrt{\epsilon-1}}\!\right)$&$\frac{-3}{4 \tilde{z}^2 \tilde{\delta}} \!\Ima\! \left(\!\frac{\epsilon}{\sqrt{\epsilon-1}}\!\right)$\\
\hline
\end{tabular}
\end{center}
The integrals being continuous functions of the variables $(z,\delta)$ for $z$ and $\delta$ near $+\infty$, the limits $\tilde{z}\to+\infty$ and $\tilde{\delta}\to+\infty$ commute. In conclusion, for $z\to+\infty$ $\mathbf{B}(\omega)$ is the only relevant contribution governing the interplay between the slab and environmental temperatures in the long-distance atomic dynamics.

We now focus on the properties of $\mathbf{D}(\omega)$ for small atom-slab distances $z\to 0$. As we will see, this evanescent contribution is divergent in this limit and is at the origin of the fact that the effective temperature at any frequency $\omega$ and thickness $\delta$ tends to the one of the slab for small $z$. Indeed, the TE components are finite because of the following behavior for $s\to+\infty$: $\Ima\bigl(\rho_1(s,\delta,\epsilon)\bigr) \sim \Ima(\epsilon)/4s^2$. The divergence comes instead from the TM components. In order to obtain the divergent TM small-distance behavior we use the following asymptotic theorem \cite{Dieu}: let $h$ be a function such that $h(s)$ has a polynomial behavior near $s=+\infty$, i.e. $h(s)\sim{\cal{H}}\,s^n$ for some non-negative integer $n$. Then when $\tilde{z}\to0$, it holds that
\begin{equation}\int_0^{+\infty} e^{-2\tilde{z}s}h(s)\,ds\sim{\cal{H}}\frac{n!}{2^{n+1}}\frac{1}{\tilde{z}^{n+1}}.\end{equation}
The quantity $\Ima\bigl(\rho_2(s,\delta,z)\bigr)$ has expansions around $s=+\infty$ given by
\begin{equation}\Ima\bigl(\rho_2(s,\delta,\epsilon)\bigr) \sim\Ima\Bigl(\frac{\epsilon-1}{\epsilon+1}\Bigr).\end{equation}
We then obtain
\begin{center}\begin{tabular}{|c|c|c|c|c|}
\hline
& $D_{xy/z}^{(\text{TE})}$ & $D_{x y}^{(\text{TM})}$ & $D_{z}^{(\text{TM})}$\\
\hline
\rule[-4mm]{0mm}{1cm}
\!\!$z\to0$ & No divergence & $\frac{3\Ima\bigl(\frac{\epsilon-1}{\epsilon+1}\bigr)}{16\tilde{z}^3}$ & $\frac{3\Ima\bigl(\frac{\epsilon-1}{\epsilon+1}\bigr)}{8\tilde{z}^3}$\\
\hline\end{tabular}\end{center}
We observe that in this limit there is no dependence on slab thickness $\delta$.

The integrands being not continuous near $(z,\delta)\sim (0,0)$  the limits $z\to 0$ and $\delta\to 0$ do not commute (the notation $\delta \fl a,\, z \fl b$ means that first the limit $z \fl b$ is taken, and then the limit $\delta \fl a$ is taken) and we have

\begin{center}
\begin{tabular}{|c|c|c|c|}
 \hline
& $D_z^{(TM)}$ & $D_{xy}^{(TM)}$ & $D_{xy}^{(TE)}$ \\
 \hline
$\delta \fl 0,\, z \fl 0$ &
$\frac{3}{8 } \frac{1}{\tilde{z}^3}\, I_1$ &
$\frac{3}{16 } \frac{1}{\tilde{z}^3}\, I_1$ &
$\frac{3  \tilde{\delta}}{2} \int_0^{+\infty} \Ima(\frac{r_1}{1-r_1^2}) ds$ \\
\hline
$z \fl 0,\, \delta \fl 0$ &
$\frac{3}{64 }\,\frac{\tilde{\delta}}{\tilde{z}^3}\, I_2$ &
$\frac{3}{128 }\,\frac{\tilde{\delta}}{\tilde{z}^3}\, I_2$ &
$\frac{3 \tilde{\delta}}{2} \int_0^{+\infty} \Ima(\frac{r_1}{1-r_1^2}) ds$ \\
 \hline
 $\delta \fl +\infty,\, z \fl 0$ &
$\frac{3}{8 } \frac{1}{\tilde{z}^3}\, I_1$ &
$\frac{3}{16 } \frac{1}{\tilde{z}^3}\, I_1$ &
$\frac{3}{4} \int_0^{+\infty} \Ima(r_1) ds$ \\
 \hline
 $z \fl 0,\, \delta \fl +\infty$ &
$\frac{3}{8 } \frac{1}{\tilde{z}^3}\, I_1$ &
$\frac{3}{16 } \frac{1}{\tilde{z}^3}\,I_1$ &
$\frac{3}{4} \int_0^{+\infty} \Ima(r_1) ds$ \\
 \hline
\end{tabular}
\end{center}
where $I_1=\Ima \left(\frac{\e-1}{\e+1}\right)$ and $I_2=\Ima \left(\frac{\e^2-1}{\e}\right)$.

R. Messina, M. Antezza, and P. Ben-Abdallah, Phys. Rev. Lett. 109, 244302 (2012).Ó


\begin{thebibliography}{99}
\bibitem{NotaTau}We remark here that in the expression of the transmission coefficient of Eq. (93) of Ref. \cite{MesAntPRA11} a factor $e^{-ik_z\delta}$ is missing. This does not modify the physical discussion provided in the context of the numerical anlysis.
\end{thebibliography}
\end{document}